\documentclass[iop, numberedappendix, twocolappendix, appendixfloats, apj, tighten]{emulateapj}
\usepackage{graphicx}
\usepackage[space]{grffile}
\usepackage{latexsym}
\usepackage{textcomp}
\usepackage{longtable}
\usepackage{multirow,booktabs}
\usepackage{amsfonts,amsmath,amssymb}
\usepackage{natbib}
\usepackage{url}
\usepackage{hyperref}
\usepackage{url}
\usepackage{listings}
\usepackage{appendix}
\usepackage{bold-extra}
\slugcomment{Submitted to The Astrophysical Journal (ApJ)}
\newcommand{\superscript}[1]{$^{\mathrm{#1}}$}

\shorttitle{Trident Method Paper}
\shortauthors{Hummels, Smith, \& Silvia}

\begin{document}

\title{Trident: a universal tool for generating synthetic absorption spectra\\ from astrophysical simulations}


%\author{Cameron~B.~Hummels\altaffilmark{1,2}, Britton~D.~Smith\altaffilmark{3}, and Devin~W.~Silvia\altaffilmark{4,2}}
%
%\altaffiltext{1}{TAPIR, California Institute of Technology, Pasadena, CA 91125, USA}
%\altaffiltext{2}{National Science Foundation Astronomy and Astrophysics Postdoctoral Fellow}
%\altaffiltext{3}{San Diego Supercomputer Center, University of California, San Diego, CA 92121, USA}
%\altaffiltext{4}{Department of Physics and Astronomy, Michigan State University, East Lansing, MI 48824, USA}

\author{Cameron~B.~Hummels\altaffilmark{1,$\dagger$}}
\author{Britton~D.~Smith\altaffilmark{2}}
\author{Devin~W.~Silvia\altaffilmark{3,$\dagger$}}
\affil{\superscript{1}TAPIR, California Institute of Technology, Pasadena, CA 91125, USA}
\affil{\superscript{2}San Diego Supercomputer Center, University of California, San Diego, CA 92121, USA}
\affil{\superscript{3}Department of Physics and Astronomy, Michigan State University, East Lansing, MI 48824, USA}

\altaffiltext{$\dagger$}{NSF Astronomy and Astrophysics Postdoctoral Fellow}

%\email{chummels@gmail.com}

%\author{Cameron B. Hummels\altaffilmark{1}}
%\affil{TAPIR, California Institute of Technology, Pasadena, CA 91125, USA}

%\author{Britton D. Smith}
%\affil{San Diego Supercomputer Center, University of California, San Diego, CA 92121, USA}

%\and

%\author{Devin W. Silvia\altaffilmark{1}}
%\affil{Department of Physics and Astronomy, Michigan State University, East Lansing, MI 48824, USA}

%\altaffiltext{1}{National Science Foundation Astronomy and Astrophysics Postdoctoral Fellow}

\begin{abstract}
Hydrodynamical simulations are increasingly able to accurately model physical systems on stellar, galactic, and cosmological scales; however, the utility of these simulations is often limited by our ability to directly compare them with the datasets produced by observers: spectra, photometry, etc.  To address this problem, we have created \textsc{trident}, a Python-based, open-source tool for post-processing hydrodynamical simulations to produce synthetic absorption spectra and related data.  \textsc{trident} can (i) create absorption-line spectra for any trajectory through a simulated dataset mimicking both background quasar and down-the-barrel configurations; (ii) reproduce the spectral characteristics of common instruments like the Cosmic Origins Spectrograph; (iii) operate across the ultraviolet, optical and infrared using customizable absorption line lists; (iv) trace simulated physical structures directly to spectral features; (v) approximate the presence of ion species absent from the simulation outputs; (vi) generate column density maps for any ion; and (vii) provide support for all major astrophysical hydrodynamical codes.  \textsc{trident} was originally developed to aid in the interpretation of observations of the circumgalactic medium (CGM) and intergalactic medium (IGM), but it remains a general tool applicable in other contexts.
\end{abstract}


%% subject header.  Objects should be in the appropriate "individual"
%% headers (e.g. quasars: individual, stars: individual, etc.) with the
%% additional provision that the total number of headers, including each
%% individual object, not exceed six.  The \objectname{} macro, and its
%% alias \object{}, is used to mark each object.  The macro takes the object
%% name as its primary argument.  This name will appear in the paper
%% and serve as the link's anchor in the electronic edition if the name
%% is recognized by the data centers.  The macro also takes an optional
%% argument in parentheses in cases where the data center identification
%% differs from what is to be printed in the paper.

%\keywords{line: formation}

\section{Introduction}
\label{sec:intro}

\subsection{Why observe simulated data?} 
\begin{figure}
\begin{center}
\includegraphics[width=0.9\columnwidth]{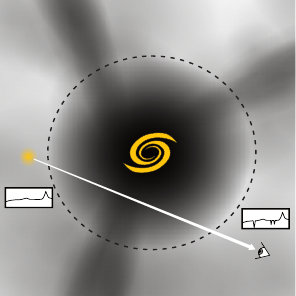}
\caption{{Schematic showing how absorption line spectroscopy indicates the presence of low-density gas through its absorption of light from background objects.  Ions in the intervening gas between an observer and a bright background object absorb discrete wavelengths of light providing information about the composition and phase of the intervening gas.  Here, an observer detects absorption from the CGM (inside dashed circle) and filamentary IGM (outside dashed circle) in the spectrum from a background quasar.
\label{fig:schematic}%
}}
\end{center}
\end{figure}

Most of the baryonic material in the universe consists of low-density
gas insufficiently bright to be detected by its emission alone
\citep[e.g.,][]{cen94, zhang95, miralda96, hernquist96, dave01}.  In order to
reveal this gas, observers rely on its ability to absorb certain
wavelengths of light from bright background sources, like how the sun
appears red to viewers looking at it through an Earth-bound dust
cloud.  The location of a background source in the sky thus defines a
sightline, usually parameterized as an infinitesimally thin
one-dimensional line, that probes the intervening material between us
and the background object.  Electron energy transitions in the
intervening gas preferentially absorb light at discrete wavelengths,
creating troughs in the spectrum of the light along this sightline.
The atoms and ions present in the intervening gas determine the viable
electron transitions, which when coupled with the relative velocity to
the observer, produce the distribution of absorption-line features in
the observed spectrum. Thus, the characteristics of different
absorption features in a sightline's spectrum can reveal an enormous
amount of information about the density, temperature, velocity,
radiation field, and ionic composition of gas along a given line of
sight. See Figure \ref{fig:schematic} for a schematic of this
process.

``Absorption-line spectroscopy'' is employed in a variety of environments where observers attempt to detect low-density gas, ranging from the gas between stars (interstellar medium -- ISM), the gas surrounding galaxies (circumgalactic medium -- CGM), and the gas between galaxies (intergalactic medium -- IGM).  These observations provide us with clues as to (i) how galaxies balance external gas accretion \citep[e.g.][]{rubin16, lehner16} with turbulent outflows of material from supernovae \citep[e.g.][]{hummels12, fielding16} and supermassive black holes \citep[e.g.][]{kacprzak15, johnson15}; (ii) what is happening in the vast volume of ``empty'' space between galaxies \citep[e.g.][]{peeples13}; and (iii) what can the CGM/IGM tell us about the evolution of the universe as a whole \citep[e.g.][]{mcquinn16}.

Observers require that background sources used in absorption-line spectroscopy be sufficiently bright and well-characterized, so that the baseline spectra are constrained when identifying absorption features due to intervening material.  Typically, quasars are used as background sources, but because bright quasars are relatively uncommon, few cosmic structures such as galaxies or gas filaments can be probed by multiple sightlines.  Thus in order to study these gas structures, observers must combine samples of sightlines through many different galaxies, making assumptions about the homogeneity of the probed galaxy population in order reach general conclusions (e.g. \textsc{cos-halos} -- \citealp{tumlinson13}; \textsc{cos-dwarfs} -- \citealp{bordoloi14}; \textsc{KBSS} -- \citealp{rudie12},  \citealp{steidel10}, \citealp{chen10}, \citealp{prochaska11}, \citealp{nielsen13}, \citealp{liang14}, \citealp{rubin14}, \citealp{turner15}).  These samples contain an enormous amount of information about the galaxy population, but as of yet, the details of the relationship between the CGM and host galaxy are not well understood.

In the last decade, significant advances have occurred in the field of hydrodynamical modeling of astrophysical systems.  There now exist many high-resolution  simulations that track the positions and velocities of stars as well as the phase and composition of gas through galaxies and significant cosmological volumes (e.g. \textsc{eris} -- \citealp{guedes11}; \textsc{fire} -- \citealp{hopkins14}; \textsc{illustris} -- \citealp{vogelsberger14}; \textsc{eagle} -- \citealp{schaye15}).  These simulations possess sufficient detail to follow the distribution and evolution of individual gas, metal, and ionic species self-consistently, making them ideal aids in understanding what is truly happening in the low-density gas so difficult to track observationally.  

The best way to compare observations and simulations is to directly
compare similar data products.  The production of synthetic observations of simulated datasets
enables such a comparison by modeling the way light
travels through space and into telescopes.  Astrophysics
has a rich history in producing mock spectra in many contexts
including stars \citep[e.g.][]{kurucz79}, the Sun \citep[e.g.][]{husser13}, galaxies
(e.g. \textsc{starburst99} -- \citealp{leitherer99}), stellar
population synthesis (e.g. \textsc{fsps} -- \citealp{conroy09}), molecular clouds
(e.g. \textsc{radmc-3d} -- \citealp{dullemond12}), plasmas
(e.g. \textsc{cloudy} -- \citealp{ferland98}), and dust
(e.g. \textsc{dusty} -- \citealp{nenkova00}).  Additionally, there
exist a number of open-source, Monte Carlo radiative transfer codes
(e.g. \textsc{sunrise} -- \citealp{jonsson06}; \textsc{hyperion} --
\citealp{robitaille11}) that are potentially applicable to
synthetic absorption-line spectroscopy, but they currently lack line transfer
physics necessary to produce absorption features.

A number of works have made use of specialized tools for generating
synthetic spectra from simulation data, each designed with the
specific features and needs of their own data formats in mind.
Examples include \textsc{specexbin} \citep{oppenheimer06} and
\textsc{specwizard} \citep{schaye03} for \textsc{gadget}
\citep{springel01,springel05}; the method of \citet{shen13} for
\textsc{gasoline} \citep{wadsley04}; the tools of \citet{churchill15}
and \citet{liang16} for their specific versions of \textsc{art}
\citep{kravtsov97,kravtsov99}; the methods of \citet{smith11} and
\citet{hummels13} (early versions of the work presented here) for
\textsc{enzo} \citep{bryan14}; and the
\textsc{fake\_spectra}\footnote{\url{https://github.com/sbird/fake\_spectra}}
code \citep{bird15} for \textsc{arepo} \citep{springel10}.  These
tools span a range of simulation methodologies, from adaptive
mesh-refinement (AMR) to smoothed-particle hydrodynamics (SPH) and
moving mesh techniques, and so must work with fundamentally different
quantities.

Thus, there is a need for a universal tool for generating mock
absorption-line spectra, one that can work with many different
simulation code formats.  Like a real telescope facility, this virtual
telescope is most beneficial when it is publicly available, not used
solely by its designers and their collaborators.  Such a publicly
available, universal tool prevents unnecessary duplication of efforts,
provides a single resource for members of the scientific community to
contribute their specific strengths, ensures fewer bugs and more
features, and enables inter-code simulation comparison efforts 
(e.g. \textsc{agora} project -- \citealp{kim14}).

\subsection{Introducing \textsc{trident}}
\label{sec:trident}
\textbf{This paper announces the full public release of
\textsc{trident}, an open-source tool for generating synthetic
absorption spectra from astrophysical hydrodynamical simulations.}
\textsc{trident} is an object-oriented, pure Python library with
support for both Python 2 and Python 3.
\textsc{trident} relies on the ability of the 
\textsc{yt}\footnote{\url{http://yt-project.org}} analysis toolkit 
\citep{turk11} to ingest simulation data from a vast array of sources
and formats for further analysis and processing.  As a result,
\textsc{trident} is capable of generating spectra for at least 10
different simulation codes.  In addition to spectral creation,
\textsc{trident} provides other analysis tools, such
as a method for creating fields of ion densities (e.g., \ion{C}{4}, \ion{O}{6}, etc.)
from simulation data using various photo- and collisional ionization
models.  \textsc{trident}
can also operate in parallel using the message passing interface
system \citep[\textsc{mpi},][]{mpi} to scale to
many processors and speed up execution (See Appendix \ref{app:parallel} for details).

Members of the scientific community are actively encouraged to use and develop \textsc{trident} as a community code\footnote{\url{http://trident-project.org}}\footnote{\url{https://doi.org/10.5281/zenodo.821220}} according to the Revised BSD License.
Table \ref{tab:resources} lists locations for various important resources
related to \textsc{trident} including its documentation, mailing list, and 
source code repository.

This paper is composed as follows.  Section \ref{sec:method} describes
the three-pronged approach \textsc{trident} takes to generating
spectra: creating ion fields for the simulation dataset
(Section \ref{sec:ion-balance}), making sightlines through the dataset
to sample the relevant fields (Section \ref{sec:sightline}), and
depositing absorption lines based on the characteristics of gas
 along the sightline (section \ref{sec:spectrum}).  Section
\ref{sec:demonstration} provides an annotated Python script
demonstrating the use of \textsc{trident} to create a simple spectrum. 
 In Section \ref{sec:discussion},
we perform some tests of \textsc{trident} including a comparison with
observational data and a curve-of-growth analysis, and we describe 
some of \textsc{trident}'s assumptions and limitations.  Finally, Section \ref{sec:summary}
summarizes the highlights and features of the code.

\begin{table}[ht!]
\begin{center}
\label{tab:resources}
\begin{tabular}{ c  c} \\
Resource & Location\\
\hline 
Web Page & {\scriptsize \url{http://trident-project.org}}\\
Source Code & {\scriptsize \url{https://github.com/trident-project/trident}}\\
Documentation & {\scriptsize \url{http://trident.readthedocs.org}}\\
Mailing List & {\scriptsize trident-project-users@googlegroups.com} \\
\hline
\end{tabular} 
\caption{Important Trident Resources
\label{tab:resources}
}
\end{center}
\end{table}

\section{Code Methodology}
\label{sec:method}

This section describes the algorithms employed by \textsc{trident} to
post-process simulation outputs.  It covers a brief discussion of:  what
simulation outputs contain and how \textsc{trident} interacts with
them (Section \ref{sec:yt}); how \textsc{trident} estimates the concentration of a desired
ion when absent from the simulation (Section \ref{sec:ion-balance}); how \textsc{trident} calculates
the trajectory of different sightlines (Section \ref{sec:sightline}); how \textsc{trident} produces
an absorption-line spectrum (Section \ref{sec:spectrum}); and how \textsc{trident} processes that
spectrum to resemble real observations (Section \ref{sec:post-processing}).  

\textsc{trident} is an object-oriented software library with its own
set of classes and modules.  The most important of these are discussed
in detail in later sections, but they are provided here as a reference:
\begin{itemize}
\item \texttt{ion\_balance} -- a module used to calculate the density of any atomic ion in a simulated dataset 
\item \texttt{LightRay} -- a class describing a one-dimensional sightline through a simulated dataset
\item \texttt{SpectrumGenerator} -- a class responsible for creating
absorption line spectra from \texttt{LightRay} objects 
\end{itemize}

For a full description of all classes and their usage in \textsc{trident}, see the API documentation\footnote{\url{http://trident.readthedocs.io/en/latest/reference.html}}.

\subsection{Brief overview of simulation outputs and how \textsc{trident} and \textsc{yt} interact with them}
\label{sec:yt}

An astrophysical hydrodynamical simulation output represents a
three-dimensional volume with a series of scalar and vector fields
expressing different fluid quantities for the gas across that volume.
Grid codes discretize the gas
into a regular grid with elements of fixed volume, oftentimes
employing AMR to achieve higher resolution in regions of interest.
SPH codes discretize the gas into particles, each representing a
parcel of gas with fixed mass, which can be smoothed using a
three-dimensional smoothing kernel to achieve a finite size.  Moving
mesh and meshless codes take a hybrid approach by representing the gas
over tessellating fluid elements that change shape as the gas moves.
Despite their differences, in all of these code formats the gas is
represented as a series of field elements describing its distribution in position,
velocity, density, temperature, metallicity, etc.

\begin{table}[ht!]
\begin{center}
\label{tab:codes}
\begin{tabular}{ c  c  c} Code & Type & Reference \\
\hline 
\textsc{arepo}\textsuperscript{a} & Moving Mesh & \citet{springel10} \\
\textsc{art-i} & AMR & \citet{kravtsov99} \\
\textsc{art-ii} & AMR & \citet{rudd08} \\
\textsc{athena} & AMR & \citet{stone08} \\
\textsc{changa} & SPH & \citet{stinson06} \\
\textsc{enzo} & AMR & \citet{bryan14} \\
\textsc{flash} & AMR & \citet{fryxell00} \\
\textsc{gadget} & SPH & \citet{springel05} \\
\textsc{gasoline} & SPH & \citet{wadsley04} \\
\textsc{gizmo} & SPH \& Meshless & \citet{hopkins15}\\
\textsc{ramses} & AMR & \citet{teyssier02} \\
\multicolumn{3}{c}{\textsuperscript{a}\footnotesize{in development}}\\
\end{tabular} 
\caption{Simulation codes that \textsc{trident} explicitly supports.
\label{tab:codes}
}
\end{center}
\end{table}

\textsc{trident} was originally developed as an analysis module within
the \textsc{yt} framework \citep{turk11}, and consequently it inherits
the way \textsc{yt} interacts with simulation outputs.  For operations
that involve sampling fluid values at arbitrary locations, particle
fields must be converted to a grid-like structure.  \textsc{yt} first
creates an underlying octree grid structure that ensures high
resolution in regions of high particle density, and then deposits
particle-based fluid quantities on to these grid elements using the
appropriate smoothing kernel and length.  These steps ensure that
subsequent analysis can be treated homogeneously across different
simulation formats and methods. Note that some operations like
calculating ion densities are performed on the particles prior to the
smoothing process.

Because of \textsc{trident}'s close relationship to \textsc{yt}, many of the features that \textsc{trident} provides to users, such as the ability to post-process a dataset to include density fields for a desired ion, can be further used within the \textsc{yt} framework seamlessly.  This enables users to make volumetric projections and slices of these ion fields, create phase plots and probability distribution functions for the presence of arbitrary ions, categorize how fluid quantities change along a line of sight, and more.

In addition, \textsc{trident} inherits \textsc{yt}'s support for every
major astrophysical hydrodynamical code.  Table \ref{tab:codes} contains a list of the simulation
codes that have been tested and confirmed to work with \textsc{trident}.
\textsc{trident} should function correctly with other
\textsc{yt}-supported codes, but appropriate sample datasets
were unavailable for testing.

\begin{figure}[ht!]
\begin{center}
\includegraphics[width=1.0\columnwidth]{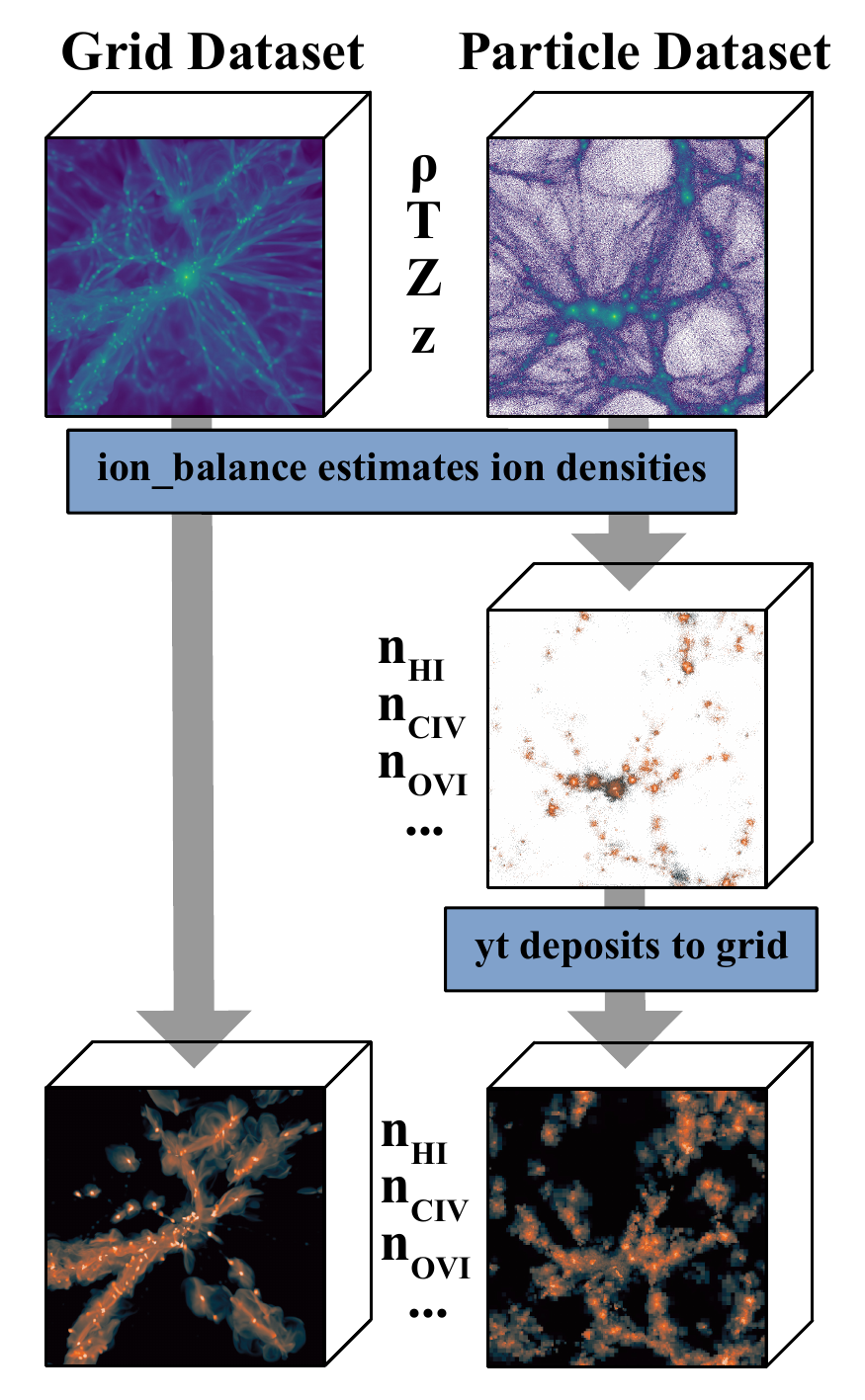}
\caption{Operation of \textsc{trident}'s \texttt{ion\_balance} module.
  Grid and particle datasets pass their redshift and fluid quantities
  (density, temperature, metallicity) to the \texttt{ion\_balance}
  module to estimate the number density of any ions of interest
  (e.g. \ion{H}{1}, \ion{C}{4}, \ion{O}{6}).  These number densities
  are computed on the original fluid elements, whether grid cell or
  particle.  Additionally, particle-based datasets are deposited onto
  a grid as an AMR octree using the particle smoothing kernel.}
\label{fig:ion_balance}
\end{center}
\end{figure}

\subsection{Creating ion fields with the \texttt{ion\_balance} module}
\label{sec:ion-balance}

In order to create absorption lines for a given ion, it is necessary that a fluid field representing the density of that ion be present for all computational elements sampled by the sightline or \texttt{LightRay}.  In some cases, these fields may be explicitly tracked by the simulation with a non-equilibrium chemistry solver.  Examples of this include \cite{smith11} and \cite{hummels13}, which follow atomic species of H and He in non-equilibrium, and \cite{cen06}, which additionally follows \ion{O}{5} through \ion{O}{9}.  However, in most cases, tracking additional ion densities within a simulation is computationally prohibitive, and so they must be derived from the available data fields using models that assume ionization equilibrium.

For species not followed by the simulation, \textsc{trident}'s \texttt{ion\_balance} sub-package creates a new field by defining the density of ion, $i$,  of element, $X$ as
\begin{equation}
n_{X_{i}} = n_{X}\ f_{X_{i}},
\end{equation}
where $n_{X}$ is the total number density of the element and $f_{X_{i}}$ is the ionization fraction of the $i$'th ion.  For simulations that track multiple metal fields, such as those presented by \cite{hopkins14a}, $n_{X}$ may already exist in the simulation output.  If this is not the case, then \texttt{ion\_balance} defines $n_{X}$ as
\begin{equation}
n_{X} = n_{H}\ Z\ \left(\frac{n_{X}}{n_{H}}\right)_{\odot},
\end{equation}
where $n_{H}$ is the total hydrogen number density, $Z$ is the metallicity and $(n_{X}/n_{H})_{\odot}$ is the solar abundance by number.  If the simulation does not explicitly track the hydrogen number density, we assume it to be given as
\begin{equation}
n_{H} = \chi \frac{\rho}{m_{H}},
\end{equation}
where $\rho$ is the total gas density and $\chi$ is the primordial H mass fraction, for which we adopt a value of 0.76. If desired, \texttt{ion\_balance} can overwrite a dataset's existing ion fields using the \texttt{force\_override} keyword to ensure consistency in ionization calculations.

\begin{figure*}
\begin{center}
\includegraphics[width=1.6\columnwidth]{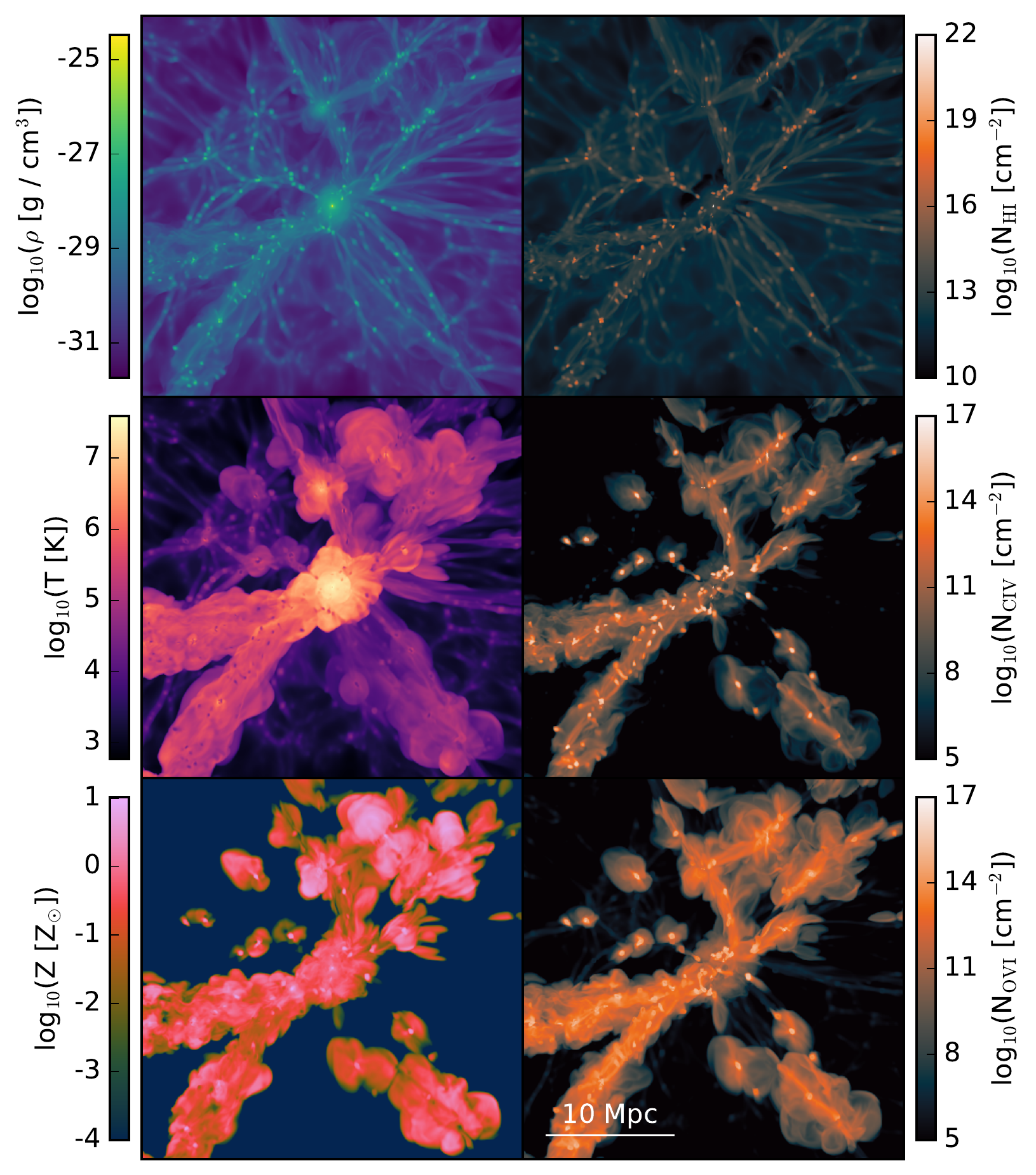}
\caption{Projections of a 1536$^3$ cosmological \textsc{enzo} dataset aimed at probing the nature of the intergalactic medium, similar to those presented by \cite{smith11}.  Each projection shows a region of the box that is 30 Mpc on a side and 5 Mpc deep. The region is centered on a halo in the simulation that was identified using the \textsc{rockstar} halo-finder \cite{behroozi12}.  The halo has a mass of $1.9~\times~10^{14}~\rm M_{\odot}$ and a virial radius of $1.5$ Mpc.  From top to bottom, the left panels show density, temperature, and metallicity, while the right panels show the projected number densities (effective column densities) of \ion{H}{1}, \ion{C}{4}, and \ion{O}{6}.  The ion number densities were computed using the \texttt{ion\_balance} module in \textsc{trident}.}
\label{ion_projections}%
\end{center}
\end{figure*}

Under the assumption of ionization equilibrium, the ionization fraction is a function of temperature, density, and the shape and intensity of the incident radiation field.  Currently, \texttt{ion\_balance} only considers radiation from metagalactic UV background models, such as those of \cite{CAFG09} and \cite{haardt12}.  In these models, the radiation field is parameterized solely by the value of the redshift, making the ionization fraction a function of just temperature, density, and redshift.  Using the code\footnote{\url{https://github.com/brittonsmith/cloudy\_cooling\_tools}.} in \cite{smith08}, we have computed the equilibrium ionization fractions for all the ions and elements through atomic number 30 (i.e., Zn) over a grid of temperature, hydrogen number density, and redshift.  These data are generated with a series of single zone simulations using the photo-ionization software, \textsc{cloudy}\footnote{\url{http://nublado.org/}} \citep{ferland13}, following the same method used by the \textsc{grackle} chemistry and cooling library \citep{smith16}.  The resulting data are saved as a three-dimensional lookup table, which is loaded by \textsc{trident} when needed.  Ionization fractions for each ion are then calculated for each computational element by linearly interpolating over these pre-computed tables.  Currently, \textsc{trident} provides data tables for the UV backgrounds described by \citet{CAFG09} and \citet{haardt12}, but the method is general enough that other backgrounds can be added.  For more information demonstrating the accuracy of these tables, see Appendix \ref{app:ion-error}.

Figure \ref{fig:ion_balance} presents a visual illustration of the way in which the \texttt{ion\_balance} module generates ion fields for simulated datasets. \texttt{ion\_balance} can create ion density fields for all computational elements for which density and temperature fields exist.  This allows fields created by \texttt{ion\_balance} to be used independently of spectrum generation, for example, to study the spatial distribution of various ions and their relationship to physical gas quantities, as shown in Figure \ref{ion_projections}.  When creating light rays from grid-based simulation data, the creation of the ion fields is saved for after the light ray is generated, allowing the fields to only be created for the cells in the light ray itself and not the entire simulation domain.  Since the three-dimensional interpolation can be computationally expensive when performed for all grid cells in the domain, this results in a significant speedup.  However, for particle-based datasets, where particle fields are smoothed onto a grid, we do not take this shortcut.  In this case, we first create the ion fields for each particle on the particle itself, and afterward deposit the resulting ion densities onto the corresponding grid cells according to the chosen smoothing kernel.  While more time consuming, this avoids errors that may arise by creating ion  fields from smoothed density and temperature fields.  

\subsection{Sightline creation: the \texttt{LightRay} object}
\label{sec:sightline}

The next step in the process of generating a spectrum is choosing and sampling a line of sight through the simulation data.  The user specifies the trajectory of the sightline through the simulation output, as well as the gas fields they wish to sample.  Optionally, the user can specify which spectral lines or ionic species it wishes to include in any subsequently generated spectrum, and \textsc{trident} will include the necessary fields.  The end product of this step is a \texttt{LightRay} object, a set of spatially-ordered, one-dimensional arrays sampling the the desired fields of the simulation output along the ray's path.  The \texttt{LightRay} is saved to an \textsc{hdf5} file \citep{hdf5} that can be reloaded by \texttt{yt} as an ordinary dataset for the purposes of spectrum generation or direct access for further analysis.

These ``light rays" make use of \texttt{yt}'s ray data container, which takes a start and end point and returns field values for all computational elements intersected by the ray's trajectory through the dataset.  For grid-based simulation codes, these computational elements are simply the highest-resolution grid cells of the Eulerian mesh along the line of sight.  As described in Section \ref{sec:yt}, particle-based codes have their particles deposited to a grid by first smoothing the particles into an octree mesh to create gridded, Eulerian fluid fields.  In this case, the computational elements returned by the LightRay are these octree cells.  

In addition to sampling the fields specified by the user, the \texttt{LightRay} creates some special fields for further processing.  For each line element along the \texttt{LightRay}, \textsc{trident} calculates and records its: \texttt{dl} --  path length; \texttt{dredshift} -- cosmological redshift interval; \texttt{redshift} -- cosmological redshift; \texttt{velocity\_los} --  line of sight velocity; \texttt{redshift\_dopp} -- doppler redshift; and \texttt{redshift\_eff} -- effective redshift.  Here we derive all of these quantities.

\begin{figure*}[ht!]
\begin{center}
\includegraphics[width=2\columnwidth]{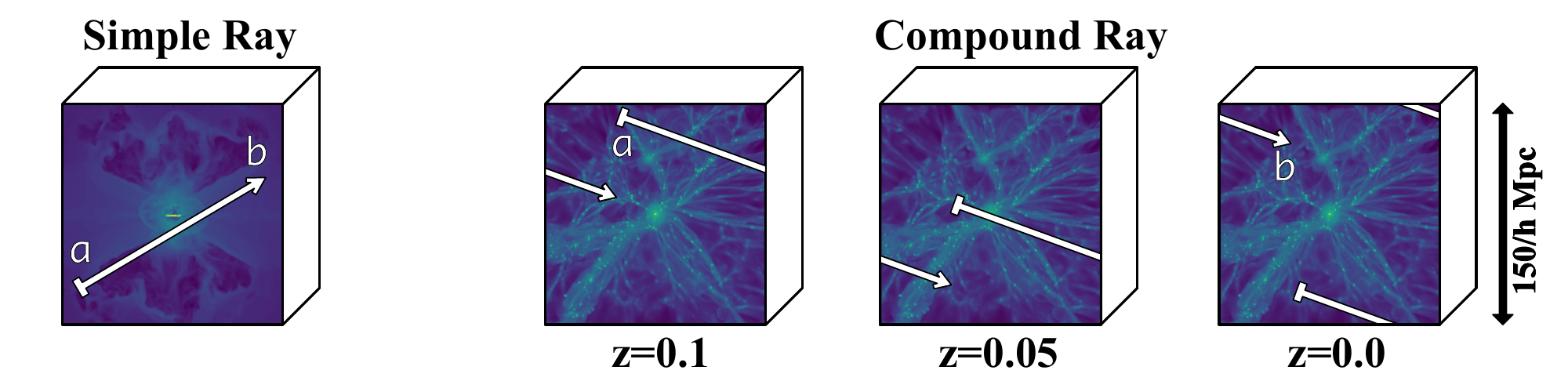}
\caption{The methods by which \texttt{LightRay} objects are generated to represent a sightline path from point $a$ to point $b$ (the observer).  \textbf{Left}: Simple rays are defined by a start point and end point for a single dataset output.  \textbf{Right}: Compound rays contain path lengths longer than the width of a single dataset by continuing the ray path over periodic boundary conditions and consecutive outputs.}
\label{fig:ray_generation}
\end{center}
\end{figure*}

Let us define the sightline of a \texttt{LightRay} vector $\vec{l}$ passing through a single dataset from point $a$ to point $b$ where the observer sits at point $b$.  We can think of it as a collection of $n$ individual line elements $d \vec{l}$:

\begin{equation}
\vec{l} = \vec{r}_b - \vec{r}_a = \sum_{i=0}^{n}{d\vec{l}_{i}}
\end{equation}

Field values along the light ray are the values within the grid or
octree cell intersected by the ray's trajectory.  The path length,
$d\vec{l}_i$, is the vector intersection of the ray with the cell.

\begin{equation}
dl_i = |d\vec{l}_i|
\end{equation}

We assume a smooth Hubble expansion between points $a$ and $b$ in the ray
such that their separation in redshift is given by the comoving radial
distance \citep{hogg99}, given by

\begin{equation}
\label{eqn:comoving-dist}
l  = D_H \int_{z_a}^{z_b} \frac{dz'}{E(z')},
\end{equation}

where

\begin{equation}
E(z) = \sqrt{\Omega_M(1+z)^3 + \Omega_k(1+z)^2 + \Omega_\Lambda},
\end{equation}

$D_H$ is the Hubble distance, and $\Omega_M$, $\Omega_k$, and
$\Omega_\Lambda$ are the ratios of mass density, spatial curvature
density, and vacuum density, respectively, to the critical density of
the universe.  Since the simulation data provide $l$, the comoving radial distance between $a$ and $b$,
and we require $z_b - z_b$, the redshift difference between $a$ and $b$, 
we must invert equation \ref{eqn:comoving-dist}.  Since there is no simple analytical
form, we use Newton's method to iteratively calculate the redshift
interval between the points, taking the redshift of the simulation
as the redshift at the \emph{far} point $a$.

Now that we know the redshift interval connecting points $a$ and $b$ within our simulation output, we assume a line element's cosmological redshift interval is linearly related to its path length:

\begin{equation}
\label{eqn:dredshift}
dz_i = \frac{dl_i}{l} (z_b - z_a)
\end{equation}

This redshift interval estimate is a good approximation on the small path lengths within an individual dataset.   This also allows us to calculate the cosmological redshift for the $i$'th \texttt{LightRay} element as:

\begin{equation}
z_i = z_b + \sum_{j=i}^{n}dz_j
\end{equation}

When generating spectra, the observer cannot discern between absorber redshifts due to cosmological expansion and its relative motion along the line of sight.  Thus, by default, \textsc{trident} also includes the effects of doppler redshift due to radial motions of the gas.  The line of sight velocity for a \texttt{LightRay} element is defined as:

\begin{equation}
v_{LOS,i} = \vec{v}_i \cdot d\vec{l}_i = v_i dl_i \cos(\theta_i)
\end{equation}

where $v_i$ is the local gas velocity field in the cell and $\theta_i$ is the angle between the line of sight and the gas velocity vector.  The local velocity field enables us to calculate a \texttt{LightRay} element's doppler redshift $z_{dopp,i}$ as

\begin{equation}
1 + z_{dopp,i} = \frac{1 + \frac{v_i}{c}\ cos(\theta_i)}
{\sqrt{1 - \left(\frac{v_i}{c}\right)^{2}}},
\end{equation}

where $c$ is the speed of light.  The \emph{effective} redshift, the redshift used to modify the location of spectral absorption lines, for each element of the ray is then given as a combination of its cosmological and doppler redshifts \citep{peebles93} as:

\begin{equation}
1 + z_{eff,i} = (1 + z_{dopp,i})\ (1 + z_i).
\end{equation}

\texttt{LightRay} generation is divided into two use cases, simple rays and compound rays.  Aside from the differences in generating their trajectories, these objects have similar internal structures and are treated the same by the rest of the \textsc{trident} machinery.

\subsubsection{Simple rays: rays traversing a single dataset}
\label{sec:simple-ray}
Simple rays are defined for use with a single simulation output at a fixed point in time.  The primary use case for simple rays is the creation of spectra from targeted physical structures, such as a specific galaxy at a particular redshift.  To generate a simple ray, the user must specify the simulation output dataset and the starting and ending locations of the ray in the dataset volume.  \textsc{trident} uses the simulation dataset's redshift as the redshift at the back of the ray (location $a$), and increments it forward along its path to the user according to the method described in equations \ref{eqn:comoving-dist}-\ref{eqn:dredshift}.  For non-cosmological simulation outputs, \textsc{trident} defaults to using a redshift of zero, but the user can specify any desired value.  Figure \ref{fig:ray_generation} illustrates a simple ray object traversing a simulation dataset.  

\begin{figure}[ht!]
\begin{center}
\includegraphics[width=1.0\columnwidth]{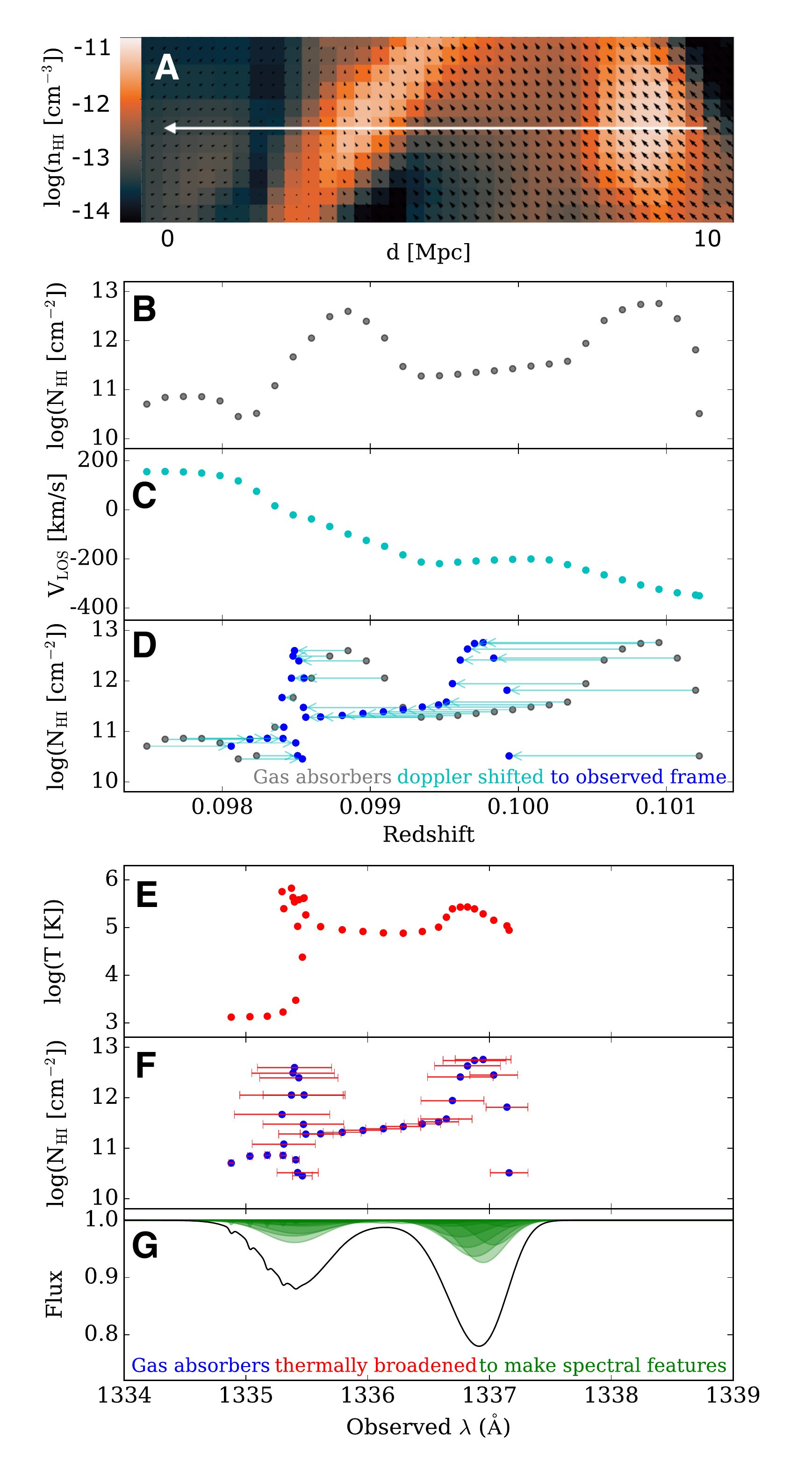}
\caption{Physical structures traced to spectral features as \textsc{trident} generates a spectrum.  Begin at the top and travel downward \textbf{A-G}, noting vertical alignment between plots. \textbf{A}: A 2D slice in neutral hydrogen density taken from a low-resolution  simulation of some cosmological filaments.  Black arrows represent the gas velocities, and a white arrow indicates the 10 Mpc-long \texttt{LightRay} sightline passing through the slice to the observer on the left.  \textbf{B}: Individual points represent gas cells probed by the sightline, plotted as cosmological redshift versus neutral hydrogen column density.  \textbf{C}: Line of sight velocity for each gas cell along the \texttt{LightRay}.   \textbf{D}:  Applying a doppler shift from the line of sight velocity shifts the location of each gas absorber into \emph{effective} redshift, the frame of the observer. \textbf{E}: In this new effective redshift frame, equivalent to $(1+z) 1216 \AA$ for the Lyman-$\alpha$ transition, the temperature of each gas cell is plotted versus observed wavelength. \textbf{F}: Points indicate the \ion{H}{1} column density in this effective redshift frame with thermal widths determined from gas temperature denoted as red error bars.   \textbf{G}: Voigt profiles calculated for each gas cell for the Lyman-$\alpha$ transition in green, superimposing to the black profile, the spectral feature seen by the observer.}
\label{fig:ray-plot}
\end{center}
\end{figure}

\subsubsection{Compound rays: rays traversing multiple datasets}
\label{sec:compound-rays}
Synthetic spectra that resemble those arising from real QSO sight lines require light rays that are many times longer than the box size of typical cosmological simulations.  For example, a comoving radial distance of 150 Mpc/h at z = 0 corresponds to a change in redshift of only $\sim0.05$.  Even with a larger box size, using a single dataset to generate a spectrum that probes the material between a distant QSO and an observer would be improper as it would fail to capture the temporal evolution of structure occurring over the light travel time within a single simulation output.

In order to create light rays spanning cosmological distances, \textsc{trident} splices together ray segments from multiple datasets written at different redshifts of the simulation.  This process was first described by \cite{smith11}.  To create these compound light rays, the user must provide the parameter file of the original simulation as well as the desired start and end redshift.  Machinery within \texttt{yt} determines the redshifts of all datasets from information stored in the simulation parameter file.  Using this layout and the framework described in equations \ref{eqn:comoving-dist}-\ref{eqn:dredshift}, we calculate the precise datasets and path length through each dataset required to span the desired redshift range.  The process for this is similar to that described in Section \ref{sec:simple-ray} in that we calculate the change in redshift equivalent to traversing the entirety of the box at the redshift of any given dataset.  The full compound ray is then constructed by piecing together the line segments from the required datasets, as illustrated in Figure \ref{fig:ray_generation}.  By default, these segments are chosen at random locations and trajectories within the box to avoid probing the same structures multiple times at different redshift.  If desired, the user has the option of maintaining a single, constant trajectory (as in Figure \ref{fig:ray_generation}), where the end point of one segment is used as the start point for the next to avoid spatial discontinuities.

\subsection{Using \texttt{SpectrumGenerator} to make spectra }
\label{sec:spectrum}

The \texttt{SpectrumGenerator} class contains all of the machinery to create absorption line spectra from the \texttt{LightRay} objects.  In order to instantiate a \texttt{SpectrumGenerator}, \textsc{trident} needs some information about the characteristics of the spectrograph modeled.  These details include the desired wavelength range, the size of individual wavelength bins, and optionally the line spread function of the spectrograph.  Users can create their own custom spectrographs or select one of the existing presets like observing mode G130M of the Cosmic Origins Spectrograph (COS) aboard the Hubble Space Telescope (HST).

In addition, users must provide the details of the absorption lines they wish to observe in their datasets.  For a given absorption line to be modeled, \textsc{trident} needs information about the corresponding quantum transition including its source ion, wavelength $\lambda$, oscillator strength f$_{\rm val}$, and probability of transition $\Gamma$.  \textsc{trident} includes a list of 220 absorption lines frequently used in CGM and IGM studies in the UV and optical (line data extracted from NIST\footnote{\url{https://www.nist.gov/pml/atomic-spectra-database}} using AstroQuery package\footnote{\url{https://astroquery.readthedocs.io}} \citep{sipocz16}), but users can easily add their own or subsample this list.

Recall that the \texttt{LightRay} object consists of a series of one-dimensional arrays of different fields (e.g. temperature, density, path length, $n_{HI}$, etc.) along its trajectory through the simulation volume.  Before creating the spectrum, \texttt{SpectrumGenerator} assures that all of the necessary ion density fields are present on the ray object needed to calculate optical depths for the desired absorption lines.  If they are not present, \texttt{ion\_balance} constructs them on the \texttt{LightRay} object itself using the gas fields.  \texttt{SpectrumGenerator} multiplies the \texttt{dl} (path length) field against each of the relevant ion number density fields to produce an array of ion column densities, corresponding to the column density of each ion for each parcel of gas intersected by the ray.  

Finally, \textsc{trident} steps through the ray object from back to 
front, depositing Voigt profiles for each encountered absorber at the appropriate wavelength for
each of the requested lines.  The wavelengths are shifted
appropriately to account for the \emph{effective} redshift of the
absorber (see Section \ref{sec:sightline}).  \textsc{trident} will add Lyman continuum absorption features for any neutral hydrogen source it
encounters, each operating as an opacity source below 912 \AA~ in the rest frame with optical depth approximated as a power law $\tau \propto \lambda^3$
\citep{rybicki79}.  Once it has passed through the entirety of the
ray, and looped over each desired absorption line, it calculates the
flux array from the optical depth array as: flux $f = e^{-\tau}$
inherently assuming that the only variations in the flux array are due
to the absorption of measured species present in the ray.

Figure \ref{fig:ray-plot} illustrates the process by which the \texttt{SpectrumGenerator} produces a Lyman-$\alpha$ absorption feature from a sightline passing through a low-resolution simulation volume.  This figure clearly depicts how physical structures can be traced directly to features in the final spectrum based on density, temperature, and velocity data.

The resulting spectrum can subsequently be post-processed to make it resemble realistic telescopic data (see Section \ref{sec:post-processing}), or it can be saved to disk as is.  \textsc{trident} supports saving spectra as tab-delimited text files, as \textsc{hdf5} files, or as \textsc{fits} files.  \textsc{trident} also contains a sophisticated plotting routine built on top of \textsc{matplotlib} \citep{matplotlib} for plotting the spectrum in various ways quickly and easily.

\subsubsection{Voigt profile calculation}
\label{sec:voigt_calc}
A spectral absorption line is caused by an atom or ion absorbing incident light of a particular energy in order to boost itself to a higher quantum energy state.  In an ideal environment with a single particle, the result is an absorption line consisting of a perfect delta function at the wavelength corresponding to the energy of the difference between the particle's quantum states.  However, in practice there are a number of processes that broaden this delta function based on the characteristics of the gas.  The two most important processes are doppler broadening, due to the velocity distribution of the gas particles, and pressure broadening, caused by the collisions of the gas particles against each other.  Doppler broadening is well-described by a Gaussian function, whereas pressure broadening can be modeled with a Lorentzian function.  The convolution of these two functions is called the Voigt profile, and it is commonly used to model spectral line profiles, yielding a value for the optical depth $\tau$ at different wavelengths.  For reference, Section \ref{sec:curve-of-growth} demonstrates the Voigt Profile shape at various spectral line strengths.

The Voigt Profile in \textsc{trident} is calculated consistent with the method described in \citet{hill16} reproduced in part here. 
The Voigt Profile $V(x, \sigma, \gamma)$ is the convolution of the Gaussian Profile $G(x, \sigma)$ and the Lorentzian Profile $L(x, \gamma)$, where $x = \nu - \nu_0$, the range of frequencies relative to the line center frequency, $\sigma$ is the standard deviation of the Gaussian, and $\gamma$ is the half-width half-max of the Lorentzian:

\begin{equation}
G(x, \sigma) = \frac{1}{\sigma \sqrt{2 \pi}} \exp\left(\frac{-x^2}{2\sigma^2}\right) 
\end{equation}

\begin{equation}
L(x, \gamma) = \frac{\gamma/\pi}{x^2+\gamma^2}
\end{equation}

\begin{equation}
V(x, \sigma, \gamma) = \int_{-\infty}^{\infty}{G(x', \sigma)L(x-x', \gamma)d\lambda'}
\end{equation}

Let us adopt $\lambda$ as our independent variable instead of $x$, since \textsc{trident} operates in wavelength space.  
The Voigt Profile possesses no closed form, but it can be numerically calculated as:
\begin{equation}
V(\lambda, \sigma, \gamma) = \frac{\Re[w(z)]}{\sigma \sqrt{2 \pi}}\textrm{, where }z = \frac{u+ia}{\sigma\sqrt{2}}
\end{equation}
and $w(z)$ is the Faddeeva function, a scaled complex complementary error function \citep{poppe90}, defined as:
\begin{equation}
w(z) = \exp{(-z^2)} \left(1 + \frac{2i}{\sqrt{\pi}}\int_{0}^{z}{e^{t^{2}}}dt\right) 
\end{equation}
The complex components of $z$ consist of $u$, our range of wavelengths relative to line center, and our damping parameter $a$:
\begin{equation}
u = c\left(\frac{\lambda_0}{\lambda} - 1\right)\textrm{ and }a = \frac{\Gamma\lambda_0}{4\pi}
\end{equation}

where $c$ is the speed of light, $\lambda_0$ is the central wavelength of the Voigt Profile, and $\Gamma$ is the sum of the transition probabilities (i.e., Einstein A coefficients) for the ionic transition.

This optical depth, $\tau$, for a line is calculated by scaling the resulting Voigt Profile by the peak optical depth $\tau_0$ at the spectral line's center \citep{armstrong67}:

\begin{equation}
\tau(\lambda, \sigma, \gamma) = \tau_0 V(\lambda, \sigma, \gamma)
\end{equation}

\begin{equation}
\tau_0 = \frac{\pi e^2 N f_{val} \lambda_0}{m_e c}
\end{equation}

where $e$ and $m_e$ are the charge and mass of the electron, $c$ is the speed of light, $N$ is the absorber's column density, and $f_{val}$ is the oscillator strength of the ionic transition.

\begin{figure}[t!]
\begin{center}
\includegraphics[width=1.0\columnwidth]{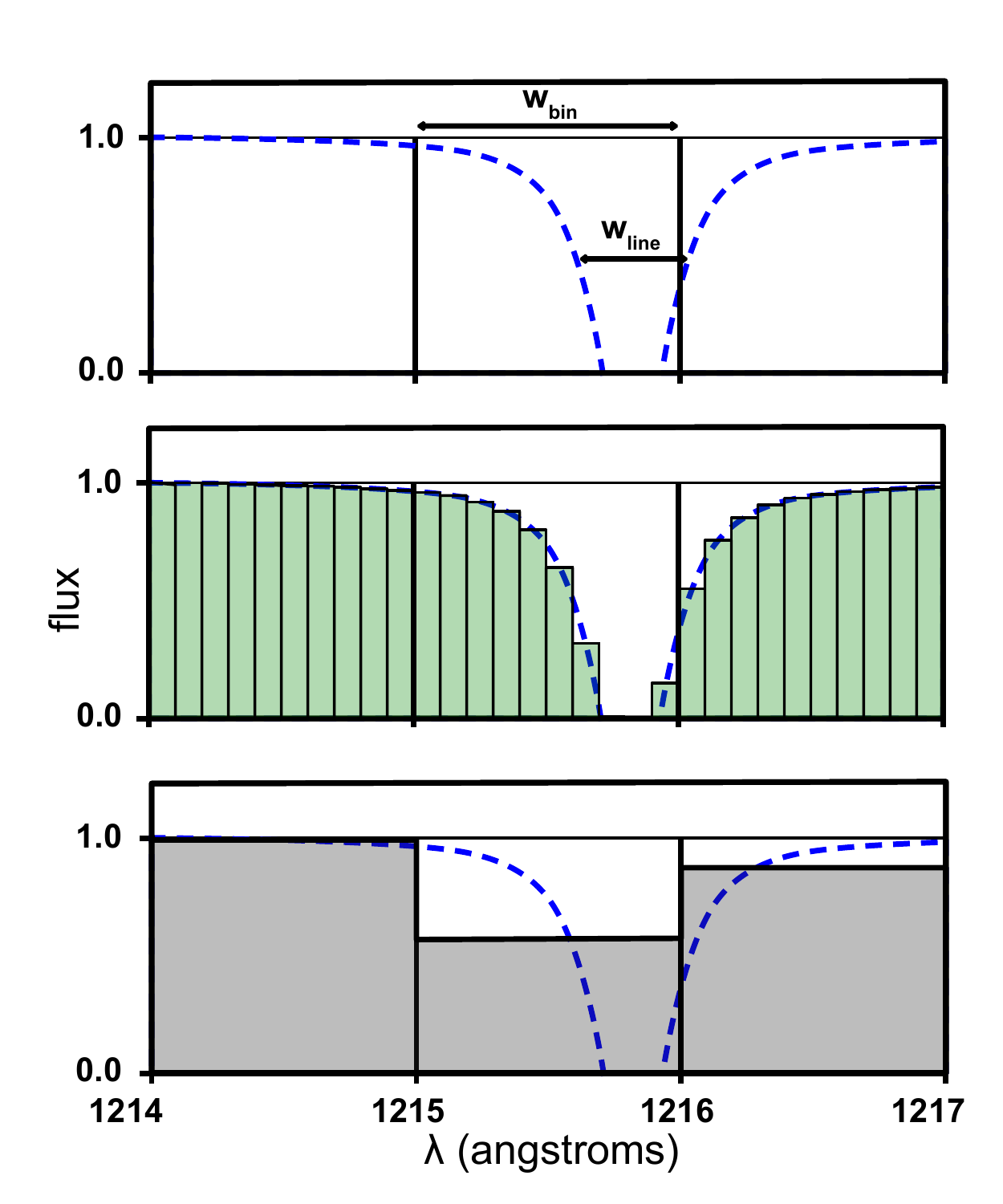}
\caption{Subgrid deposition method.  When absorption features are narrower than the spectral bin width, \textsc{trident} creates an array of virtual bins at higher spectral resolution to assure flux conservation of unresolved features.  \textbf{Top}: Depositing a narrow Lyman-$\alpha$ feature (dashed blue line) into a spectrum with coarse 1 angstrom spectral bins  $w_{line} < w_{bin}$).  \textbf{Middle}: \textsc{trident} creates an array of virtual bins with 0.1 angstrom resolution (green bins) into which we deposit the Voigt profile to approximate the line's flux deficit.  \textbf{Bottom}: We numerically integrate the area of the virtual bins to calculate the equivalent width of the unresolved spectral line on our original coarse bins (grey bins). }
\label{fig:subgrid_deposition}%
\end{center}
\end{figure}

\subsubsection{Voigt profile deposition} 
\label{sec:voigt_depo}
Each time a spectral feature is added, \textsc{trident} identifies the wavelength bin where its deposition will be centered.  However, because there is no closed form for the Voigt Profile, it is difficult to know \emph{a priori} how wide a spectral absorption feature will extend in wavelength space.  Therefore, \textsc{trident} adaptively increases the size of the window over which it deposits the spectral feature, sampling the Voigt Profile at the center of each bin location in the spectral window.  \textsc{trident} repeats this operation until the window is wide enough that the deposited $\tau$ values at the edges of the window are less than $10^{-3}$ before depositing the entire Voigt Profile into the spectrum.

On the other hand, spectral features that are too small, narrower than the chosen wavelength bin width, would be ignored by the algorithm and lost since the \texttt{SpectrumGenerator} only calculates the Voigt Profile at the centers of each wavelength bin. Ignoring these narrow features leads to the total $\tau$ and flux being dependent on the chosen wavelength bin width, which is inherently unphysical.

To address this problem, \textsc{trident} performs subgrid deposition when it recognizes that the thermal width of a spectral line is narrower than the spectral bin width in \texttt{SpectrumGenerator}.  Subgrid deposition creates an array of virtual spectral bins, each less than one tenth the thermal width of the spectral feature.  \textsc{trident} deposits the Voigt Profile to these virtual bins, then numerically integrates them to determine the equivalent width of the spectral line at the original low-resolution wavelength bin.  This process conserves $\tau$ and total flux regardless of the wavelength bin width used in the output spectrum.  The process of subgrid deposition is illustrated in Figure \ref{fig:subgrid_deposition}.

\begin{figure*}[ht!]
\begin{center}
\includegraphics[width=2\columnwidth]{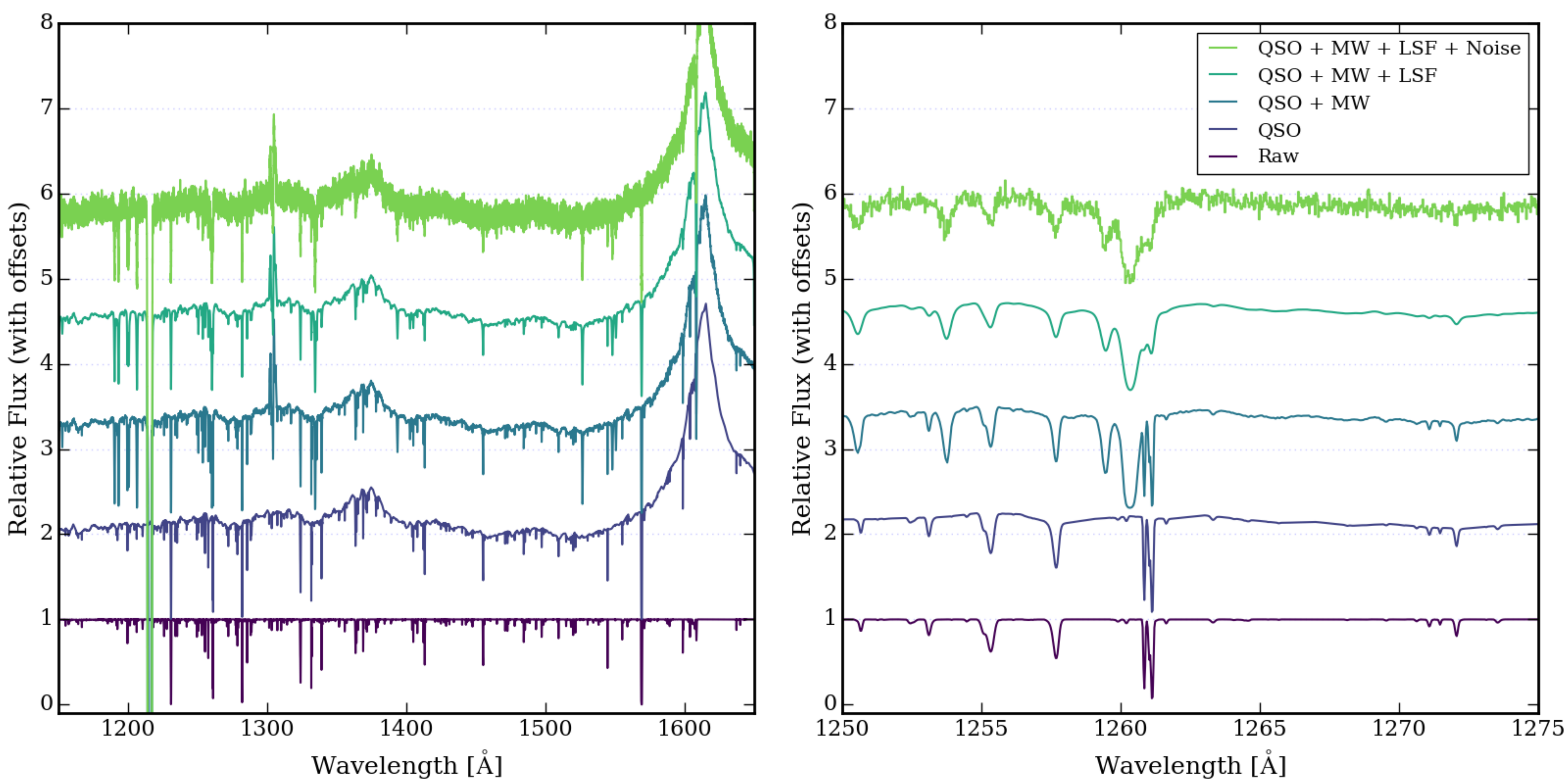}
\caption{From bottom to top, we show the progression of an increasingly complex synthetic absorption spectrum from a clean, raw spectrum to one with all of the components described in Section \ref{sec:post-processing}. The spectral components are added in the following order: Composite quasar spectrum (QSO), Milky Way foreground (MW), line spread function (LSF), noise. The left panel shows the entirety of the generated spectrum while the right panel shows a zoomed in region of the spectrum to allow one to see the significance of the post-processed modifications of the original spectrum. The light ray used for this spectrum comes from a 1536$^3$ cosmological \textsc{enzo} dataset aimed at probing the nature of the intergalactic medium and traverses a range in redshift from $z = 0.0$ to $z = 0.03295$. It includes a set of common UV spectral lines.}
\label{spectrum_progression}%
\end{center}
\end{figure*}

\subsection{Post-processing the spectrum}
\label{sec:post-processing}
Once a ``raw'' spectrum has been generated, additional levels of complexity can be added by \textsc{trident}. These additional features are intended to produce progressively more realistic synthetic spectra that can be directly compared to observational datasets using the same analysis tools employed by observers.  Figure \ref{spectrum_progression} illustrates how \textsc{trident} can post-process a raw spectrum to make it more realistic according to the steps below.

First, in order to make comparisons with observational quasar sightlines, the most basic spectrum modification includes the addition of an underlying quasar spectrum at a desired redshift, usually the far redshift $z_a$ of the \texttt{LightRay}. This is accomplished by taking the composite QSO spectrum calculated by \citet{telfer02}, shifting it to the desired redshift, and computing an interpolated relative flux as a function of wavelength. This interpolated and shifted spectra is then multiplied by the raw spectrum to add the effects of the background quasar. To further approach realism with our synthetic spectrum, we can also introduce spectral features due to foreground contamination from the Milky Way (MW).  Similar to the method introduced for adding a background QSO spectrum, instead we use the average MW foreground\footnote{\url{https://archive.stsci.edu/prepds/igm/}} computed by \citet{danforth16}.

Once a spectrum is produced that contains all of the desired observational signatures, one can further modify the spectrum by applying a set of instrument-specific properties, as suggested in Section \ref{sec:spectrum}. First, to most accurately match the desired instrument, the initial spectrum uses the known pixel resolution of the instrument for the wavelength bin size then, in this post-processing step, the spectrum is convolved with the line spread function (LSF) of the specified instrument. \textsc{trident} can currently convolve TopHat and Gaussian kernels with its spectra, and it additionally accepts custom kernels.  In the case of mimicking the Cosmic Origins Spectrograph, we convolve the spectrum with an average LSF computed for each observing mode's kernel\footnote{\url{http://www.stsci.edu/hst/cos/performance/spectral\_ resolution/}}. Furthermore, \textsc{trident} readily allows for the addition of other instrument properties through its built-in \texttt{Instrument} class.  

Finally, we allow for the addition of Gaussian random noise.  For a specified value of the signal-to-noise ratio (SNR), random fluctuations drawn from a Gaussian distribution are added to the spectrum.  Alternatively, an arbitrary noise vector can be supplied by the user.  The resulting spectrum is as realistic as possible.  
\section{Demonstration: how to run \textsc{trident}}
\label{sec:demonstration}

Here we provide an annotated example Python script for a common use-case of \textsc{trident}.  This script generates a COS spectrum of a sightline passing through the center of an \textsc{art-ii} dataset.  This script and others that are similar can be found in our documentation to step you through the process with different simulation codes.% and in the source code itself in \texttt{trident/examples/working\_script.py}.

First, we load the relevant Python modules of \textsc{yt} and \textsc{trident}.  We set the dataset filename and load the dataset into \textsc{yt}.  The dataset used is the publicly available\footnote{\url{http://trident-project.org/data/sample_data}} initial output of an \textsc{art-ii} run of an isolated galaxy used in an \textsc{agora} paper \citep{kim16} assumed to be at redshift of 0.  We define the trajectory of our \texttt{LightRay} sightline to cross the full domain of our simulation, and additionally define what lines, ions, or atoms we want to include in our spectrum.  \textsc{trident} is extremely flexible in terms of what lines we can include.  Here we will include all lines produced by all ions from hydrogen and silicon, singly-ionized magnesium (\ion{Mg}{2}), and the 1335 $\rm\AA$ line from singly-ionized carbon (\ion{C}{2}).

\begin{verbatim}
import yt
import trident as tri
fn = `AGORA_LOW_000000.art'
ds = yt.load(fn)
ray_start = ds.domain_left_edge
ray_end = ds.domain_right_edge
line_list = [`H', `Si', `Mg II', `C II 1335']
\end{verbatim}

Now, we create a sightline through the dataset, using the trajectory and field requirements we defined.  We will save it to disk as \texttt{ray.h5} as well as use it locally.  Note that we have set our \texttt{ftype} keyword to \texttt{`gas'}.  This is the \textsc{yt}-based field type indicating where \textsc{trident} should do the \texttt{ion\_balance} calculations.  Here we set it to \texttt{`gas'} because this is an \textsc{art-ii} dataset and AMR codes should make the ion interpolations on the grid, denoted by \texttt{`gas'}.  However, for SPH datasets, the interpolation must occur on the particle itself before being smoothed to the grid.  Thus one would set \texttt{ftype} to be the field type associated with the frontend's gas particles (e.g. \texttt{PartType0} for \textsc{gadget} and \textsc{gizmo}, \texttt{Gas} for \textsc{gasoline}, etc.).

\begin{verbatim}
ray = tri.make_simple_ray(
                ds, 
                start_position=ray_start,
                end_position=ray_end, 
                lines=line_list, 
                ftype=`gas')
\end{verbatim}

We can then take a look at the path of our sightline through the simulated volume by using \textsc{yt}'s functionality to create an image of our simulated volume down the $x$-axis in projected gas density, zoom in on the center, overplot the path of the ray on our projection, and save it to disk.

\begin{verbatim}
p = yt.ProjectionPlot(ds, `x', `density')
p.annotate_ray(ray)
p.zoom(20)
p.save(`projection.png')
\end{verbatim}

From this \texttt{LightRay} object we just created, we will generate an absorption spectrum using the defaults associated with the Cosmic Origins Spectrograph instrument aboard the Hubble Space Telescope.  This sets things according to the G130M observing mode where the spectral range is 1150-1450 \AA, the spectral bin size is 0.01 \AA, and the appropriate line spread function is applied.  The user could easily define their own instrument with arbitrary settings.  This raw spectrum is now saved to a tab-delimited text file, and the spectrum is plotted to an image.

\begin{verbatim}
sg = tri.SpectrumGenerator(`COS')
sg.make_spectrum(ray, lines=line_list)
sg.save_spectrum(`spec_raw.txt')
sg.plot_spectrum(`spec_raw.png')
\end{verbatim}

Lastly, we do some post-processing to the resulting spectrum, adding in a background quasar, and the Milky Way foreground, applying the defined line spread function to it, and adding gaussian noise with a signal-to-noise ratio of 30.  These steps are performed to make our data as much like spectra an observer would obtain through a real spectrograph.  We then plot and save the ``final'' spectrum.

\begin{verbatim}
sg.add_qso_spectrum()
sg.add_milky_way_foreground()
sg.apply_lsf()
sg.add_gaussian_noise(30)
sg.plot_spectrum(`spec_final.png', step=True)
\end{verbatim}

\begin{figure}[h!]
\begin{center}
\includegraphics[width=1.0\columnwidth]{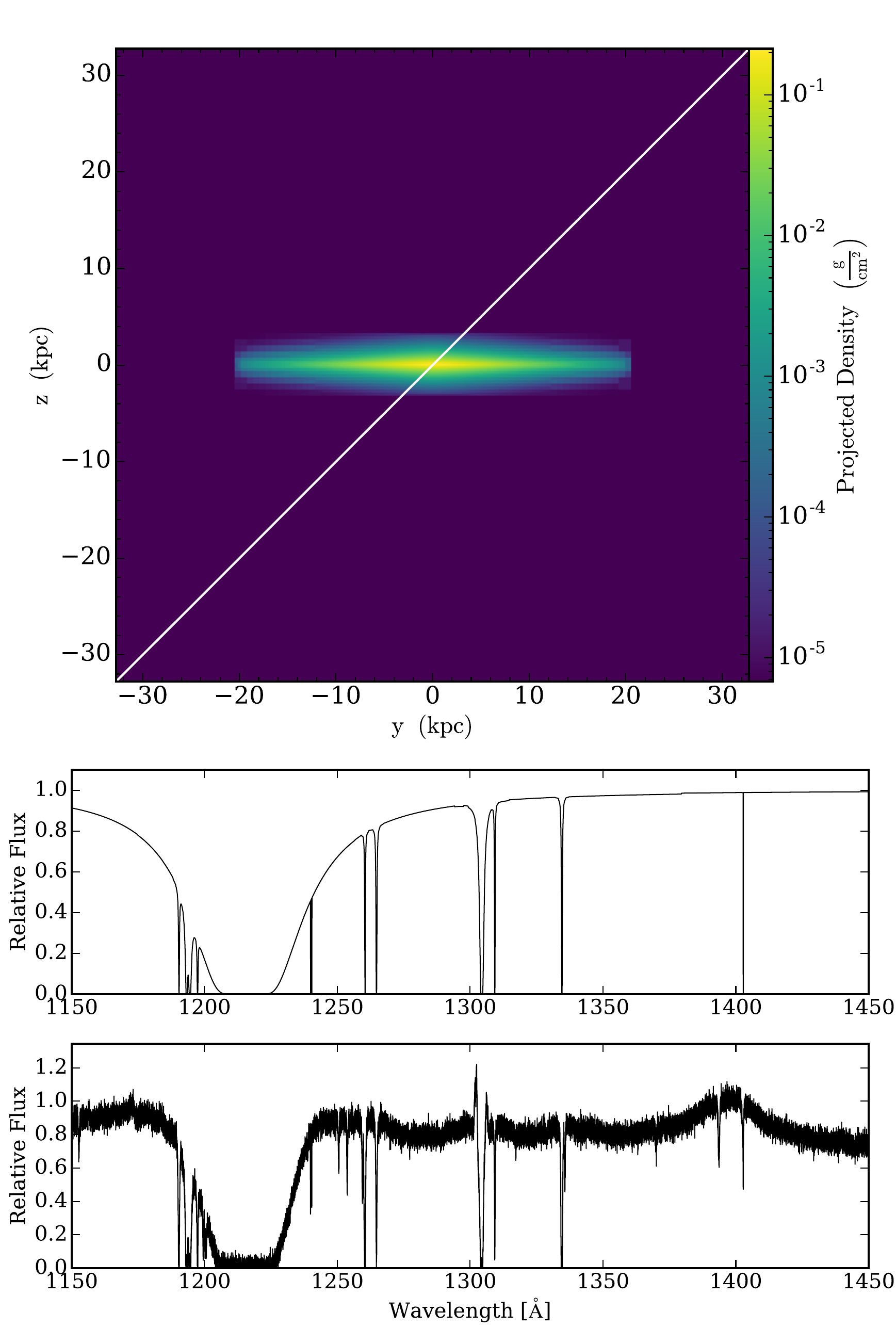}
\caption{Resulting images from our annotated demonstration.  \textbf{Top}: Density projection of isolated galaxy simulation taken from \textsc{agora} team showing path of \texttt{LightRay}. \textbf{Middle}: Raw spectrum of our \texttt{LightRay} featuring lines from H, Si, \ion{Mg}{2}, and the \ion{C}{2} 1335 \AA~line. \textbf{Bottom}: Final post-processed spectrum also including quasar background, milky way foreground, COS line-spread function, and signal to noise of 30.}
\label{fig:final_spec}%
\end{center}
\end{figure}

Figure \ref{fig:final_spec} displays the three images that are generated by this working script, showing the path of the \texttt{LightRay} sightline as it probes the isolated disk galaxy, the raw spectrum, and the post-processed COS-like spectrum.  While this dataset and script are extremely simple, we are able to reproduce the spectrum of a damped Lyman-$\alpha$ absorber with several accompanying lines including some silicon and oxygen lines.  Because this script and dataset are freely available, we encourage readers to reproduce this result on their own.

\section{Discussion}
\label{sec:discussion}

\subsection{Comparison with real spectra} \label{sec:igm}

\begin{figure}[h!]
\begin{center}
\includegraphics[width=1\columnwidth]{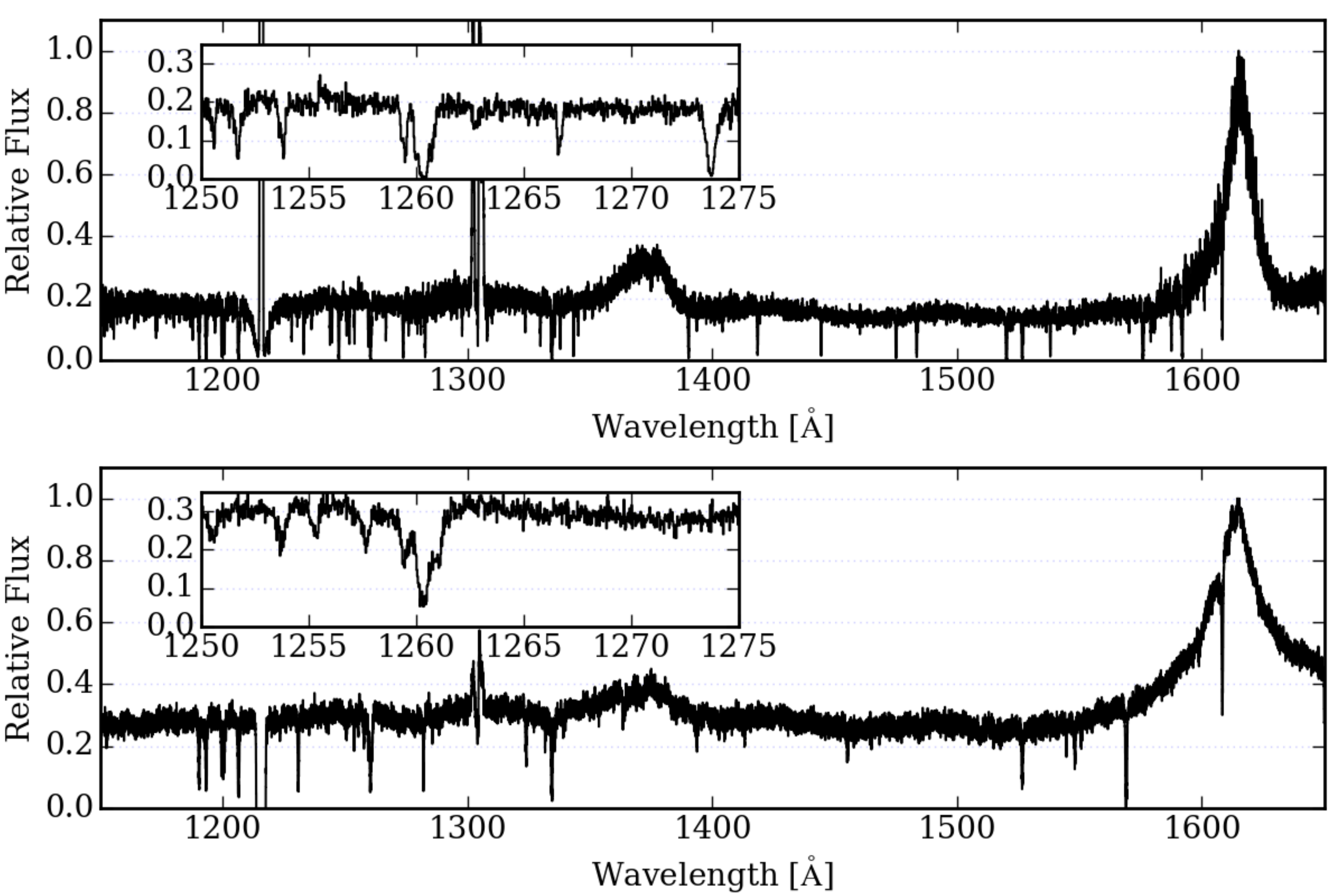}
\caption{\textbf{Top}: COS Spectrum for a QSO at z=0.3295 from data made available by \citet{danforth16}
\textbf{Bottom}: \textsc{trident} synthetic spectrum for a QSO at z=0.3295 using a light ray from an 1536$^3$ cosmological \textsc{enzo} dataset similar to those presented in \citet{smith11}. Both of the inset plots show the same zoomed-in region of wavelength space for each spectrum.}
\label{fig:igm}%
\end{center}
\end{figure}

We can assess how well \textsc{trident} creates synthetic spectra by making a direct comparison against equivalent observational data. 
IGM \texttt{LightRay}s are generally compound rays, sightlines that
pass through several simulation outputs to create a long enough
trajectory to reach high redshift sources.  Figure \ref{fig:igm}
compares a publicly-available QSO spectra from \citet{danforth16} to a
synthetic spectrum from outputs of simulation similar to those of
\citet{smith11}.  For the purposes of this comparison, the synthetic
spectrum is generated using the instrument properties of COS to match
the observational characteristics of the true spectrum.  As can be
seen, the spectra are similar in shape, in the location of major
features, and in the locations of many spectral lines.  Subsequent analysis can be performed using automated Voigt profile fitting algorithms \citep[e.g.][]{dave97, egan14} to extract the ``observed" properties of the the simulated IGM.  Taken a step further, \textsc{trident}-generated spectra like the one in Figure \ref{fig:igm}, readily enable the creation of community tools like the MAST Interface to Synthetic Telescopes with yt \citep[MISTY;][]{peeples14}, a public simulation-to-archive pipeline that simplifies the process of accessing and interacting with synthetic spectra.

\subsection{Code test: curve of growth}
\label{sec:curve-of-growth}
Due to the extreme conditions found in the low-density astrophysical environments probed by absorption-line spectra, it is very challenging to make explicit tests to assure that \textsc{trident} exactly reproduces the spectra from experimental data.  However, one viable test is to demonstrate how well \textsc{trident} reproduces the so-called \emph{curve of growth}, the relationship between the an ion's column density and its resulting absorption strength.  The curve of growth is a well-studied problem with a clear empirical-derived physical solution relating the equivalent width $W$ of a spectral absorption line to the number density $N$ of that ion along the probed sightline.  

\begin{figure}[t!]
\begin{center}
\includegraphics[width=1\columnwidth]{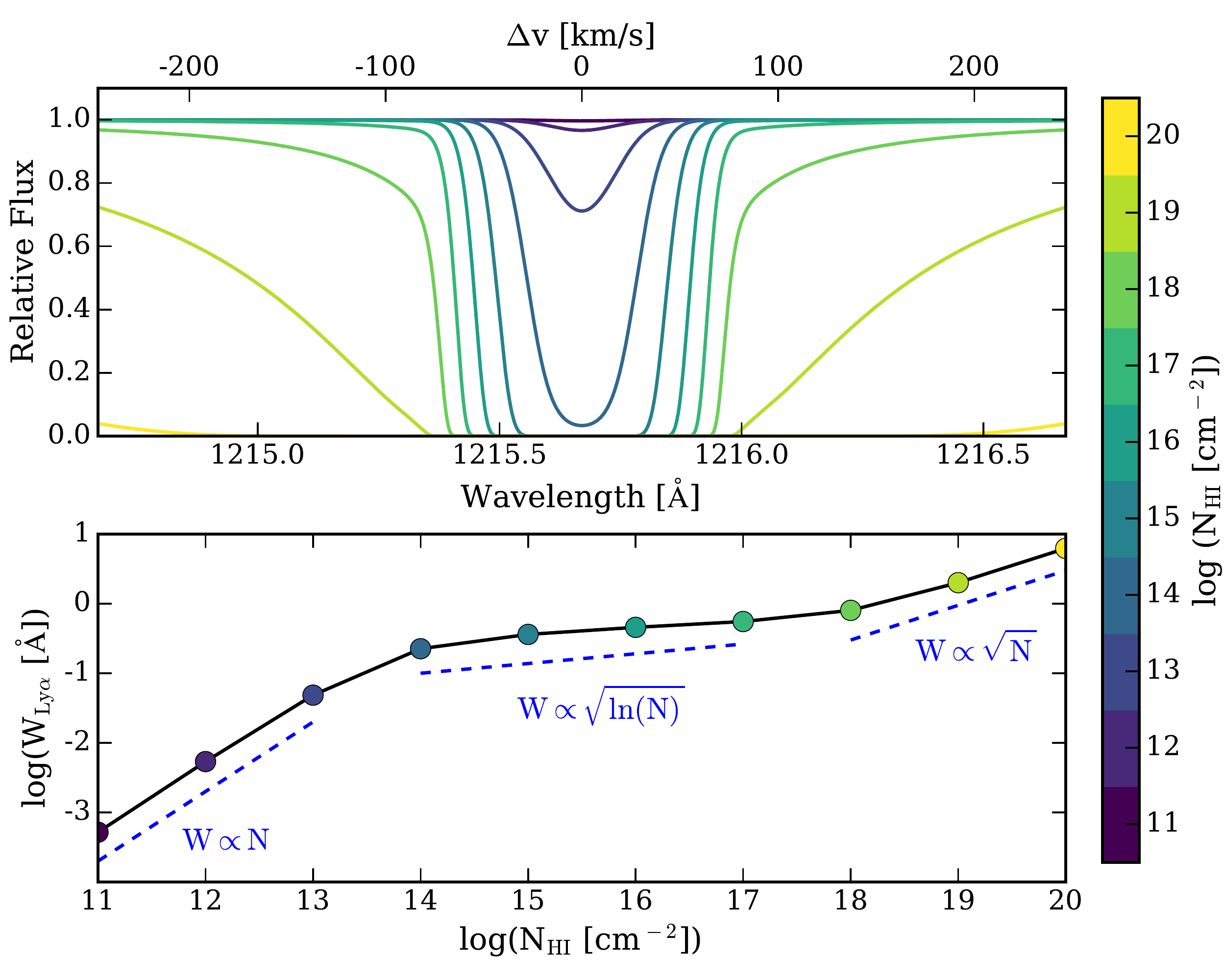}
\caption{\textbf{Top}: Voigt profiles of the Lyman-$\alpha$ line for different neutral hydrogen column densities at $T = 30,000$ K ($b = 22$ km s$^{-1}$) as produced by \textsc{trident}.  \textbf{Bottom}: Corresponding curve of growth describing the relationship between each line's column density and line equivalent width.  Dashed lines indicate the linear (left), saturated (center), and logarithmic (right) portions of the curve of growth as reproduced by \textsc{trident}.}
\label{fig:curve_of_growth}%
\end{center}
\end{figure}

As described in Section \ref{sec:voigt_calc}, absorption line shapes follow a Voigt Profile, a combination of the relatively narrow Gaussian profile with the relatively wide Lorentzian profile.  The complex shape of the Voigt profile leads to a non-linear relationship between the column density of an absorber and its corresponding spectral line strength.  Observers commonly use the equivalent width of a spectral line $W = \int{(1 - f_{\lambda} / f_0) d\lambda}$ as a proxy for its strength. In Figure \ref{fig:curve_of_growth}, we have plotted several Lyman-$\alpha$ absorption features deposited at various neutral hydrogen column densities at a doppler parameter of 22 km s$^{-1}$ on top, and the resulting curve of growth indicating their corresponding equivalent widths on bottom. 

In the optically thin limit, when an absorption line does not block out all of the flux at a given wavelength, it is said to be in the ``weak" or ``linear" regime of the curve of growth.  The Voigt Profile approximates a gaussian, where increases in column density cause proportional increases in the equivalent width of the absorption line.   The first three shallow absorption features ($N_{\rm HI} = 10^{11} - 10^{13}$ cm$^{-2}$) all sit within the ``linear" regime of the curve of growth where $W \propto N$.

As the absorption features increase in strength and become opaque enough to \emph{saturate} and block out all flux at a given wavelength, increases in column density are met with negligible increases in the line's equivalent width.  Because the Voigt Profile is still dominated by the Gaussian Profile, increases in column density do little to increase the depth of the line.  This regime is termed ``flat" or ``saturated", and it shows up in Figure \ref{fig:curve_of_growth} as the middle four absorption features ($N_{\rm HI} = 10^{14} - 10^{17}$ cm$^{-2}$) where $W \propto \sqrt{\ln{N}}$.

Finally, very strong lines start to behave more like a Lorentzian Profile, where the wings block increasing amounts of flux in surrounding wavelengths.  This regime is referred to as ``strong" or ``damped".  The growth of the equivalent width is slow in this regime, and only increases with the square root of the column density.  The ``damped" lines are the three strongest lines in Figure \ref{fig:curve_of_growth} for absorption features ($N_{\rm HI} = 10^{18} - 10^{20}$ cm$^{-2}$) where $W \propto \sqrt{N}$.

The behavior of the curve of growth in \textsc{trident} perfectly reproduces the textbook case, thereby validating these aspects of \textsc{trident}'s operation.

\subsection{Limitations of \textsc{trident} and its data tables}
\label{sec:limitations}

\subsubsection{Limitations of \textsc{trident}}
\label{sec:trident_limitations}
While the availability of a code like \textsc{trident} is a benefit to the astrophysical community, it has several limitations that should be noted when using it and interpreting its results for scientific research.  Many of the limitations listed below can be addressed in future versions of the code or by including more detail in the underlying simulations themselves before using \textsc{trident}.

\emph{\textsc{trident} does not perform full radiative transfer on the simulation.}  A radiative transfer (RT) code approximates how electromagnetic waves propagate between all of the emitters and absorbers in a simulated volume over various wavelength photons.  RT codes can produce very realistic photometry, spectroscopy, and IFU data, but they are computationally expensive to run at comparable resolution to observational data, and oftentimes they lack relevant physics (e.g. line transfer).  \textsc{trident} only tracks the absorption effects along the desired sightline, but it is fast and possesses a number of additional features missing from most RT codes making it a good complement to RT analyses.

\emph{SPH codes deposit ion fields to a grid before sightline integration.}  Because \textsc{trident} operates as an extension of \textsc{yt}, it inherits \textsc{yt}'s treatment of particle-based codes.  At present, \textsc{yt} converts particle-based codes outputs into a grid-based format in order to leverage \textsc{yt}'s extensive framework for processing and analyzing grid data.  The process is performed conservatively, depositing the particles to an adaptive grid using a \emph{scatter} operation at the cell centers to preserve the inherent dynamic resolution of the dataset and utilize the unique smoothing kernel of the original simulation code for the deposition.  Furthermore, the ionic abundance calculations of \texttt{ion\_balance} (see Section \ref{sec:ion-balance}) take place on the particles themselves prior to their deposition to the grid.  Most particle-based absorption-line synthetic spectral generators \citep[e.g.][]{oppenheimer06} calculate a sightline's column density by directly integrating its trajectory through the smoothing kernel of each intersected SPH particle.  Preliminary analysis indicates agreement between this traditional SPH integration method and the grid-based method adopted by \textsc{trident} to $< \sim 10$\% (Dong et al., in prep).  Therefore, the grid-based treatment of SPH particles is not a limitation per se, but it requires explanation.  The next version of \textsc{trident}, expected out by end of 2017, will incorporate particle kernel direct integration consistent with other particle-based spectral generation codes.

\subsubsection{Limitations of the current \texttt{ion\_balance} data tables}

It is impossible to calculate the exact abundance of a given ion by simply knowing the instantaneous gas density, temperature, and  metallicity fields, but \textsc{trident} does a fair job at estimating it.  However, there are several assumptions built into the currently available data tables used by the \texttt{ion\_balance} module to approximate ionic species abundances.  Recall that \texttt{ion\_balance} operates by plugging a gas parcel's density, temperature, and radiation field (vis-a-vis redshift) into a three-dimensional data table to interpolate and determine the relative abundance of its desired ionic species.  The lookup table is populated by thousands of individual \textsc{cloudy} runs varied over these gas densities, temperatures, and radiation fields.  For a full description of how \texttt{ion\_balance} operates, see Section \ref{sec:ion-balance} and Appendix \ref{app:ion-density}.   

\textsc{trident} provides a few data tables assuming different UV background models \citep{haardt12, CAFG09}, but at present these suffer from some limitations. These limitations can be avoided by simulating and tracking the ionic species in your hydrodynamical simulation that you wish to use in the generation of synthetic spectra and avoiding use of the \texttt{ion\_balance} module entirely.   Furthermore, users can generate their own data tables with publicly available code\footnote{https://github.com/brittonsmith/cloudy\_cooling\_tools} \citep{smith08}.  However, take heed of the following limitations based on the assumptions inherent in the default lookup tables when using \texttt{ion\_balance} to approximate ionic abundances.

\emph{Current \texttt{ion\_balance} data tables assume UV background radiation operates in the optically thin limit.}  At present, the data tables provided to \texttt{ion\_balance} were produced by including the full effect of the UV background, ignoring any self-shielding effects of gas deeply embedded in a high-opacity envelope.  Gas with neutral hydrogen columns $N_{HI} < 10^{17.2}$ cm$^{-2}$ shield nearby gas from ionizing radiation E $<$ 13.6 eV \citep{CAFG11}.  This effect has been approximated in previous work \citep{rahmati13}, and the next version of \textsc{trident}, due out by the end of 2017, will account for it.  Currently ignoring the effects of self-shielding will artificially raise the ionization state of the various low ions present to some degree, particularly in clumped regions.

\emph{Current \texttt{ion\_balance} data tables ignore local photo-ionizing sources.}   Currently, the strength and the spectrum of the UV background radiation field used to generate the lookup tables does not account for additional local sources of ionizing radiation which may be present in the simulation, like AGN and massive stars.  This is a common approach used by other groups, since including additional radiation source terms (and additional dimensions) in the \texttt{ion\_balance} data table would make it extremely large.  Notably, \citet{shen13} calculated that local photo-ionizing effects from an L* galaxy with a galactocentric star formation rate of SFR = $20 M_o / yr$ were only dominant over the metagalactic UV radiation field within 45 kpc of the galactic center.  Therefore, by not accounting for local photoionizing sources in the data tables, \texttt{ion\_balance} artificially reduces the ionization state of gas in the interiors of AGN and starburst galaxies.

\emph{\texttt{ion\_balance} data tables assume ionization equilibrium.}  In the absence of fields in the original simulation that explicitly follow the evolution of a desired ionic species, it is impossible to calculate its non-equilibrium state instantaneously.  Thus, \texttt{ion\_balance} estimates an ion's abundance by using the aforementioned photoionization from a UV background coupled with collisional ionization.  Ionization equilibrium remains valid in high-density, low-temperature regime where the cooling time is short, but it breaks down in the low-density parts of the IGM \citep{cen06}.  Studies suggest that ionization fractions can vary between equilibrium and non-equilibrium treatment of high ionization species of oxygen gas in the CGM and IGM at the $\lesssim30$\% level (\citealp{cen06}, \citealp{oppenheimer16}, Silvia et al, in prep).  In these studies, the equilibrium models predict a reduced ionization state of gas for high ions in low-density environments.

\section{Summary}
\label{sec:summary}
In this paper, we have presented \textsc{trident}, a parallel, Python-based, open-source code for producing synthetic observations from astrophysical hydrodynamical simulation outputs.  Its features include:
\begin{itemize}
\item post-processing simulation outputs to include ion density fields for any desired ion based on instantaneous fluid and ionizing radiation conditions
\item creating \texttt{LightRay}s, ordered one-dimensional arrays sampling the fields in a dataset along a chosen sightline or across multiple consecutive simulation outputs to approximate a sightline spanning a large redshift interval;
\item generating a spectrum from the fluid quantities contained in a \texttt{LightRay} object and a custom list of relevant ions and spectral lines;
\item post-processing a spectrum to match the characteristics of a spectrum observed by a real spectrograph including its wavelength range, spectral resolution, line spread function, noise, etc.;
\item full support for simulations for all major astrophysical hydrodynamical code formats;
\item automatic parallelization for both sightline and spectral generation using \textsc{mpi};
\item ability to directly trace physical structures to their resulting spectral features and vice versa (see Figure \ref{fig:ray-plot});
\end{itemize}
We encourage members of the scientific community to both use and contribute to \textsc{trident}.  For more information on acquiring, installing, and using \textsc{trident}, please see Table \ref{tab:resources}.

%\acknowledgements
\section{Acknowledgements}
\label{sec:acknowledgements}
We are extremely grateful to our co-developers within the \textsc{yt} community
for their assistance in coding, reviewing, testing, and timing releases of \textsc{yt} to
match the \textsc{trident} development schedule.  In particular, we wish
to thank Matthew Turk, Nathan Goldbaum, Kacper Kowalik, and John ZuHone for going
above and beyond the requirements of a open-source software community. Special thanks go 
out to Bili Dong, Lauren Corlies, and Andrew Emerick for early code contributions.
\textsc{trident} benefited greatly from our discussions with the
aforementioned and also Molly Peeples, Brian O'Shea, X Prochaska,
Nicolas Tejos, Jess Werk, Jason Tumlinson, Nick Earl, Kate Rubin, Nicholas
Lehner, Josh Peek, Chuck Steidel, Gwen Rudie, Sean Johnson, Ben Oppenheimer, Amanda
Ford, Romeel Dave, Robert Thompson, David Weinberg, Neal Katz, Joel
Primack, Ian McGreer, Greg Bryan, Phil Hopkins, Paul Torrey, Josh
Suresh, Simeon Bird, Du\u{s}an Kere\u{s}, Claude-Andr\'{e}
Faucher-Gigu\`{e}re, Erika Hamden, and Clayton Strawn.  We are
especially grateful to Charles Danforth and Ian McGreer for the quasar and
Milky Way templates they provided. We thank Jacob
Kneibel for his work on the early stages of generating realistic
synthetic spectra, particularly in using COS LSFs.  Also
thanks goes to Ji-hoon Kim and members of the \textsc{agora} consortium
for providing datasets to test
\textsc{trident}  across many simulation formats.  Support for this work was provided
by NASA through Hubble Space Telescope Theory Grants HST-AR-13917,
HST-AR-13919, HST-AR-13261.01-A and HST-AR-14315.001-A and 
ATP grants NNX09AD80G and NNX12AC98G from the 
Space Telescope Science Institute, operated by the Association of 
Universities for Research in Astronomy, Inc., under NASA contract NAS 5-26555.
Additional support comes from the NSF through AST grant 0908819 and
the NSF Astronomy and Astrophysics Postdoctoral Fellowship program.
Computational resources for this work were provided through NSF XSEDE
grant TG-AST140018, NSF BlueWaters grants PRAC-gka and GLCPC\_jth.
This research also used resources of the National Energy Research
Scientific Computing Center (NERSC), a DOE Office of Science User
Facility supported by the Office of Science of the U.S. Department of
Energy under Contract No. DE-AC02-05CH11231.  We
acknowledge access to NERSC resources made possible by University of
California High-Performance AstroComputing Center (UC-HiPACC).

%\software{Astropy \citep{astropy}}
\section*{Software}
\textsc{hdf}\footnote{\url{https://hdfgroup.org}},
\textsc{h5py}\footnote{\url{http://h5py.org}},
\textsc{scipy}\footnote{\url{http://scipy.org}},
\textsc{numpy}\footnote{\url{http://numpy.org}} \citep{numpy},
\textsc{matplotlib}\footnote{\url{http://matplotlib.org}}
\citep{matplotlib}, \textsc{mpi} \citep{mpi},
\textsc{mpi4py}\footnote{\url{http://pythonhosted.org/mpi4py/}}
\citep{mpi4py}, \textsc{yt}\footnote{\url{http://yt-project.org}}
\citep{turk11}, and
\textsc{astropy}\footnote{\url{http://astropy.org}} \citep{astropy}.

\appendix
\begin{appendices}
\section{Parallelism and Performance}
\label{app:parallel}
The two primary classes of \textsc{trident}, \texttt{LightRay} and \texttt{SpectrumGenerator}, are both parallelized using \textsc{mpi}.  This is implemented using the \texttt{parallel\_objects} helper function from \textsc{yt}, which is itself built upon the \textsc{mpi4py} module.  The \texttt{parallel\_objects} function is a loop iterator that divides iterations between \textsc{mpi} work groups and facilitates the re-joining of results from all groups at the end of the loop.  Use of parallelism in both \textsc{yt} and \textsc{trident} scripts requires that \texttt{yt.enable\_parallelism()} be present after module imports.

The \texttt{LightRay} creation step is parallelized by splitting up the simulation datasets required for compound ray generation over the available \textsc{mpi} processes, typically in single-process work groups.  In the case where there are more available processes than datasets (such as for simple rays), multiple processes can be allocated to an instance of a single dataset, making use of \textsc{yt}'s internal parallelism to partition the work.  

The \texttt{SpectrumGenerator} is parallelized using a similar, two-layered approach.  First, the generation of a single spectrum is parallelized over the absorption lines to be deposited (i.e., Ly-$\alpha$, Ly-$\beta$, etc.).  If the number of available processes exceeds the number of lines to be deposited, then the line deposition task itself is split among available processes for depositing each absorber.  In the limit where the number of lines and/or absorbers is much greater than the number of available processes, this parallelism strategy scales well and is, in practice, limited by the speed and parallelism of the filesystem.  In addition, the \textsc{yt} \texttt{parallel\_objects} function can be used directly by the user for the embarrassingly parallel task of operating over multiple \texttt{LightRay} objects and their subsequent spectra.

\begin{figure}[h!]
\begin{center}
\includegraphics[width=1\columnwidth]{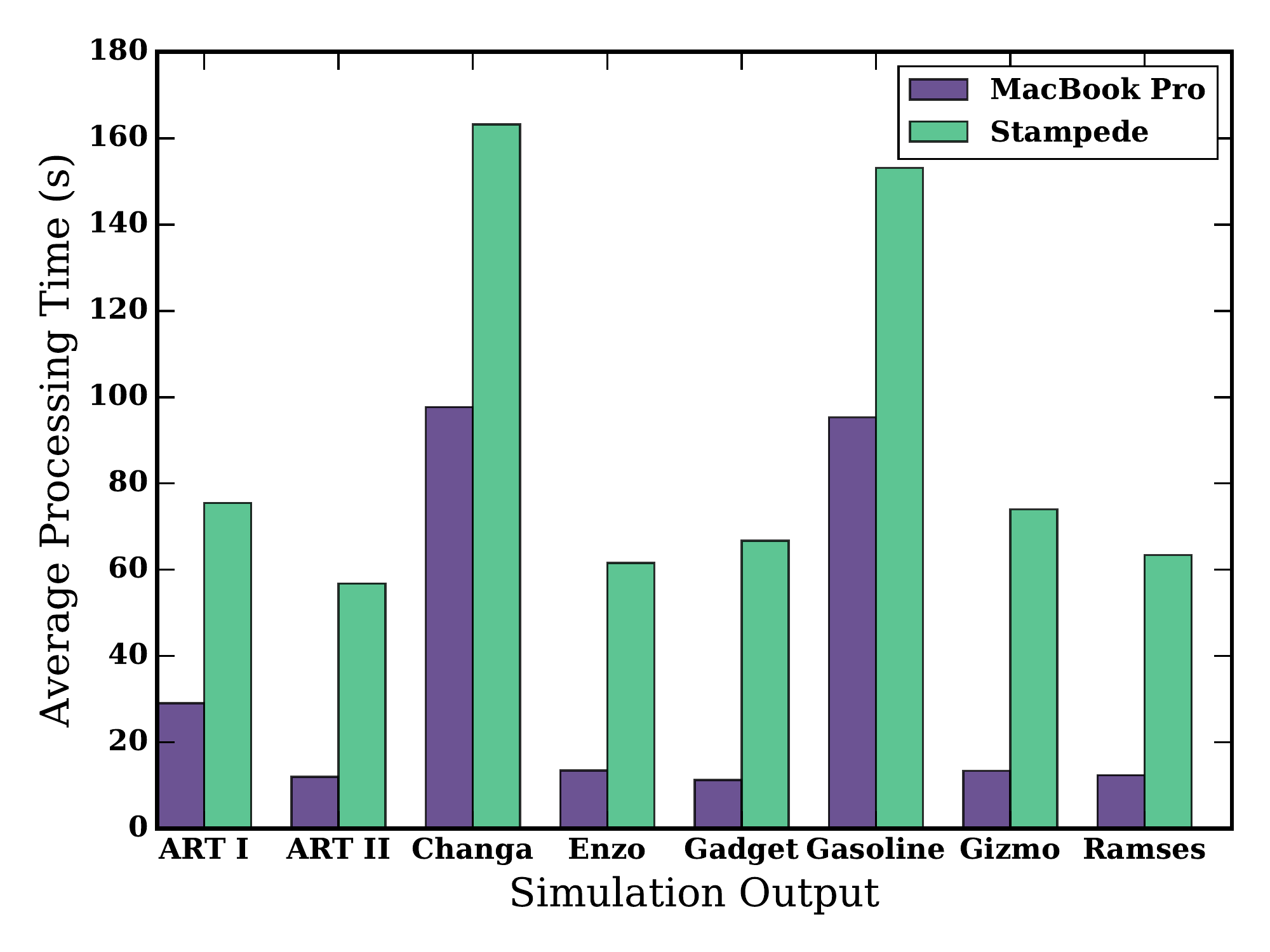}
\caption{\label{fig:performance}
Average \textsc{trident} processing time to generate a simple ray and spectrum for different simulation outputs from the \textsc{agora} isolated galaxy simulations \citep{kim16}.}
\end{center}
\end{figure}

As a reference benchmark, we ran \textsc{trident} on the \textsc{agora} idealized galaxy simulations \citep{kim16}.  We ran the script provided in Section \ref{sec:demonstration} (without the \texttt{ProjectionPlot} step) on the initial outputs for each of the simulations codes supported in the \textsc{agora} study.  These scripts were run on two machines: (1) an early 2015 MacBook Pro with 3.1 GHz Intel Core i7 processor and 16 GB RAM, and (2) a single Intel Xeon E5 Sandy Bridge processor on Stampede, the National Science Foundation's flagship supercomputing cluster run by the Texas Advanced Computing Center (TACC) as part of the Extreme Science and Engineering Discovery Environment (XSEDE) initiative.  The script was modified slightly to send ten sightlines through the central galaxy and create a spectrum for each.  Figure \ref{fig:performance} shows the average amount of time \textsc{trident} takes to generate a single sightline and spectrum.  Stampede compute cores are substantially slower than the MacBook Pro presumably due to file system load.  The increased processing time required for \textsc{gasoline} and \textsc{changa} particle-based codes reflects the particle deposition step described in Section \ref{sec:ion-balance} and the fact that fields are not cached between sightline generation.  As previously noted in Section \ref{sec:trident_limitations}, the next \textsc{trident} release will address these issues affecting particle-based code performance by numerically integrating particle kernels on the fly and avoiding the particle deposition step altogether.

Subsequent profiling of this benchmark script reveals that $\sim60\%$ of the processing time is taken by the \texttt{SpectrumGenerator}, made up of equal parts voigt profile calculation, subgrid deposition, and array bookkeeping.  Creation of the \texttt{LightRay} requires about $\sim20\%$ of the processing time primarily spent calculating what cells are intersected by the sightline and then accessing and saving these data.  The remaining time spent is miscellaneous time associated with applying the line spread function, saving the \texttt{LightRay} and spectra to disk and figures, etc.  Notably, less than 1\% of the run time is spent in the \texttt{ion\_balance} portion of \textsc{trident}.

\section{Accuracy of \texttt{ion\_balance} tables}
\label{app:ion-error}

The dependency of a given ionization state of gas is a very complex and non-linear function of its gas density, temperature, and incident radiation field.  As described in Section \ref{sec:ion-balance}, \textsc{trident} approximates this dependency by interpolating over a large three-dimensional look up table created by thousands of one-cell \textsc{cloudy} models \citep{ferland13}.  However, the resolution of this data table will determine how well this non-linear three-dimensional function can be sampled to provide adequate estimates of ionic abundances.  

We estimate the level of error for a data table by taking the sum of all ionization fractions for a given species over a grid of random points that span the density/temperature/redshift parameter-space of the data table.  In reality, all ionization fractions species for a given species should always add to 1 (e.g. $f(\rm H_{\rm I}) + f(H_{II}) = 1.0$).  We measure how much the sum of our ionization fractions for a species deviate from 1 as an estimate of the error of the data table.  

We perform this test using low and high resolution data tables for the
\cite{haardt12} UV background model used by \textsc{trident} and
available for
download\footnote{\url{http://trident-project.org/data/ion\_table/}}.
The high-resolution table (296 MB) is the default that we recommend
for users, but the low-resolution table (41 MB) is provided for users
concerned about disk space or bandwidth.  Both data tables span their
density, temperature, and redshift dimensions as: -9 $\le$ log($n_{\rm
  H}$ / cm$^{-3}$) $\le$ 4; 1 $\le$ log(T / K) $\le$ 9; and 1 $\le$
log(1 + z) $\le$ 1.2.  The low resolution table samples the density,
temperature, and redshift dimensions with 27, 161, and 22 points,
respectively.  The high resolution table samples the density,
temperature, and redshift dimensions with 105, 321, and 22 points, respectively, and is thus 8 times larger than the low resolution table.  We find the error to be significantly dependent on redshift, so we use the same set of random densities and temperatures within each redshift bin covered by the input table.

\begin{figure}[h!]
\begin{center}
\includegraphics[width=1\columnwidth]{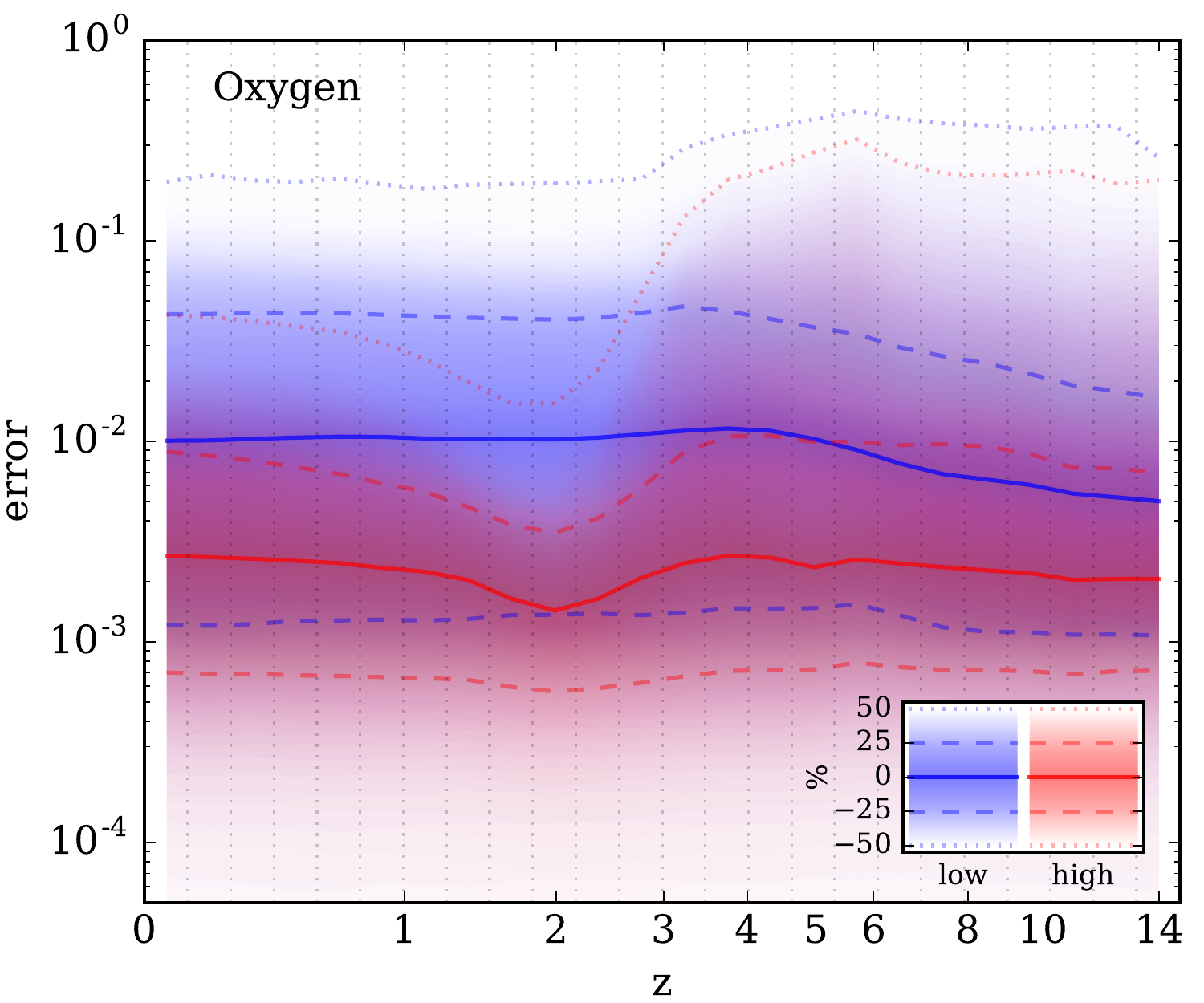}
\caption{\label{fig:error-O}
Levels of error in oxygen ionization fraction calculations by \texttt{ion\_balance} when using the Haardt-Madau data tables.  Error is defined in Appendix \ref{app:ion-error}.  Blue shaded regions show the error distribution for the low resolution table, and red shaded regions show the high resolution table.  The solid lines show the median error, dashed lines show $\pm$25\%, and the dotted line shows the maximum error.  The horizontal, grey dotted lines show the redshift bins of the input data.}
\end{center}
\end{figure}

\begin{figure}[h!]
\begin{center}
\includegraphics[width=1\columnwidth]{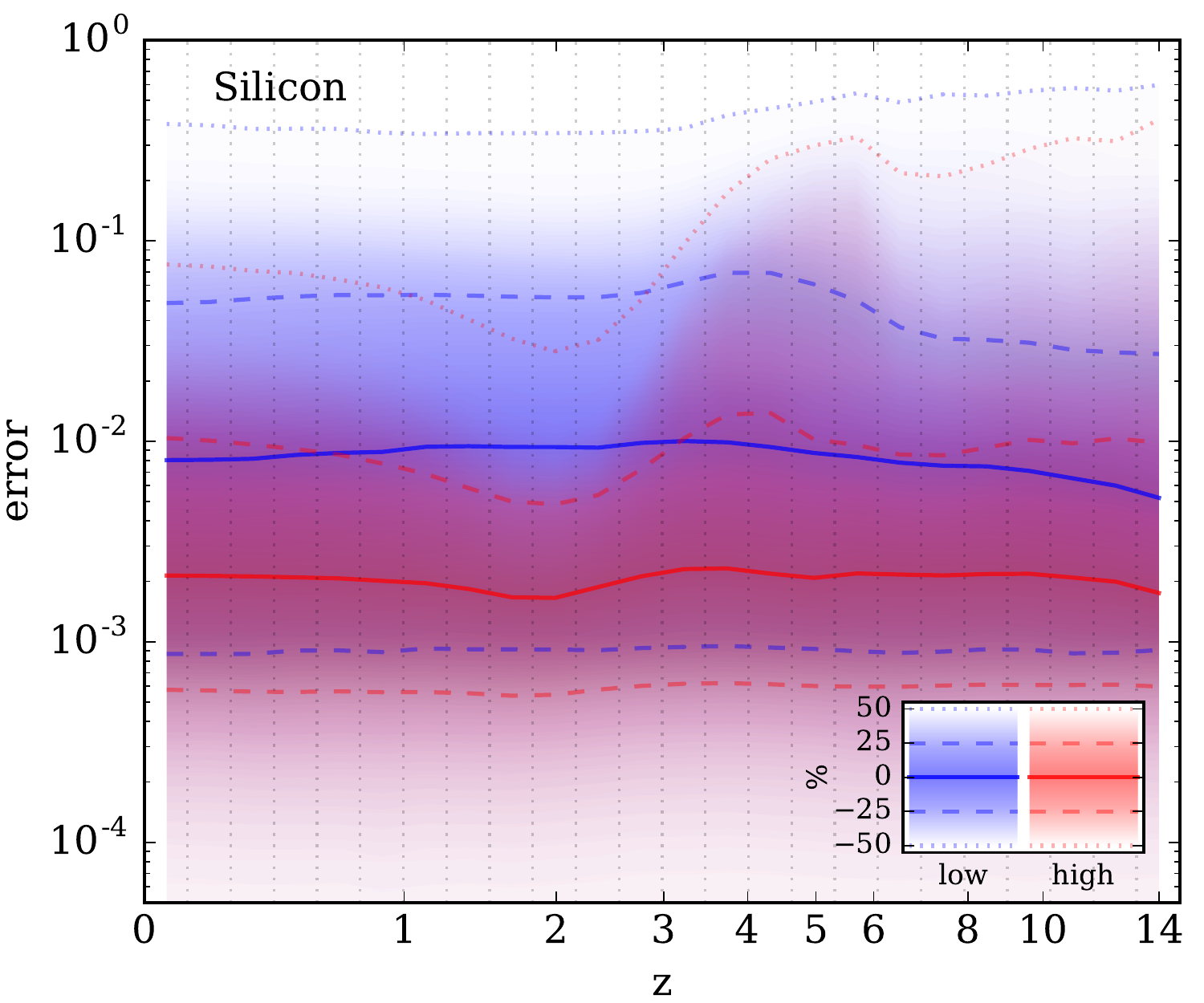}
\caption{\label{fig:error-Si}
Same as Figure \ref{fig:error-O}, but for Silicon.}
\end{center}
\end{figure}

In Figures \ref{fig:error-O} and \ref{fig:error-Si}, we show the distribution of error in ionization fraction for O and Si.  In each case, we use 100,000 random points in log($n_{\rm H}$ / cm$^{-3}$) and log(T / K) for each redshift bin.  For each redshift bin, we select random redshifts within the bin.  We define the error as
\begin{equation}
error = 1 - \sum_{i=1}^{N} f_{i},
\end{equation}
where $f_{i}$ is the ionization fraction of the i'th species of any element with N total ionization states.  We find that the total ionization fraction only ever exceeds 1 by maximally $\sim10^{-3}$, and so we only show situations where the total ionization fraction is less than 1.  For both O and Si, the average error is about 1\% for the low resolution table and about 0.2\% for the high resolution table.  We find rare cases where the error can be significantly larger.  For oxygen, the error can reach $\sim$40/32\% in the low/high resolution tables at redshifts $z \sim$ 5.5.  This is even higher for silicon, due mainly to the greater number of ionization states.  However, at redshifts less than 2, the maximal error for the high resolution table never exceeds 5\% for O and 8\% for Si.

While interpolation over the data table works well, extrapolation beyond its bounds can create some problems.  \textsc{trident} does not explicitly support data with densities or temperatures outside the range provided above.  If an extrapolation yields an unphysical ionization fraction for an ion, for instance one greater than one, \textsc{trident} will cap it at one and warn the user.  This can occur for calculating ionization fractions of low ions at exceptionally cold temperatures outside our provided temperature ranges (e.g. 1 K).

\section{Ion density generation}
\label{app:ion-density}

Here we describe the full algorithm that \textsc{trident} employs in the \texttt{ion\_balance} module to generate the number density field for a desired ionic species.  The ionic number density field can be derived in different ways depending on the fields that exist in the simulation output dataset.  In short, for a desired ion field (e.g. \ion{O}{6}), \textsc{trident} will use the on-disk field if present in the simulation, otherwise \textsc{trident} employs \texttt{ion\_balance} on the appropriate metal field when present (e.g. oxygen abundance), or in the case that the desired metal fields are not tracked at all, \textsc{trident} assumes a solar abundance of the desired metal from the bulk metallicity field (e.g. $Z$).  Solar abundance values are extracted from the documentation of Cloudy \citep{ferland98} based on previous work \citep{grevesse98, holweger01, allende01, allende02}.  Hereafter describes the full algorithm for the curious reader:

Define $X_{i}$ as the $i$'th ion of element $X$.
The total number density of $X_{i}$ is
\begin{equation}
n_{X_{i}} = f_{i}\ n_{X},
\end{equation}
where
\begin{equation}
f_{i} \equiv \frac{n_{X_{i}}}{n_{X}},
\end{equation}
and $n_{X}$ is the total nuclei number density of all species of
$X$.  In terms of the mass density, $\rho_{X}$,
\begin{equation}
\rho_{X} = n_{X}\ m_{X},
\end{equation}
where we define the atomic mass of $X$, $M_{X}$, as
\begin{equation}
m_{X} \equiv M_{X}\ m_{H},
\end{equation}
and $m_{H}$ is the Hydrogen mass.

The goal is to generate $n_{X_{i}}$, but different base fields will exist, depending on the dataset.  If $\rho_{X}$ exists,
then
\begin{equation} \label{eqn:MX1}
n_{X} = \frac{\rho_{X}}{M_{X}\ m_{H}}
\end{equation}
and
\begin{equation}
n_{X_{i}} = \frac{f_{i}\ \rho_{X}}{M_{X}\ m_{H}}.
\end{equation}

If $\rho_{X}$ does not exist, then there are two possibilities.  First, we define the solar abundance of element $X$, $A_{X}$, as
\begin{equation} \label{eqn:AX1}
A_{X} = \frac{n_X}{n_H}|_{\odot}.
\end{equation}
$X$ is either H or He.
\begin{equation}
n_{X} = A_{X}\ n_{H}.
\end{equation}
If we have $\rho_{H}$, then we can say
\begin{equation}
n_{X} = A_{X}\ \frac{\rho_{H}}{m_{H}}.
\end{equation}
If we do not have $\rho_{H}$, then we will assume the primordial H
mass fraction, $\chi=0.76$.  In that case, we have
\begin{equation}
n_{H} = \frac{\chi\ \rho}{m_{H}}
\end{equation}
and
\begin{equation} \label{eqn:nx1}
n_{X} = A_{X}\ \frac{\chi\ \rho}{m_{H}}.
\end{equation}

If $X$ is a metal, then define the metallicity of $X$ as
\begin{equation}
Z_{X} = \frac{\rho_{X}}{\rho}.
\end{equation}
If we have $Z_{X}$ and equation \ref{eqn:MX1}, then
\begin{equation}
n_{X} = \frac{Z_{X}\ \rho}{m_{H}\ M_{X}}.
\end{equation}

If we do not have $Z_{X}$, then we must use the solar abundance and the total metallicity, $Z$,
given by
\begin{equation}
Z = \frac{\rho_{metals}}{\rho},
\end{equation}
and
\begin{equation} \label{eqn:rhox1}
\rho_{X} = Z\ \rho\ \frac{\rho_{X}}{\rho}|_{\odot}
\end{equation}
Taking $\rho_{X}$ and $\rho_{H}$, we have
\begin{equation}
\frac{\rho_{X}}{\rho_{H}} = \frac{n_{X}\ M_{X}\ m_{H}}{n_{H}\ m_{H}}.
\end{equation}
Canceling out the $m_{H}$, we have
\begin{equation}
\frac{\rho_{X}}{\rho_{H}} = \frac{n_{X}\ M_{X}}{n_{H}},
\end{equation}
and
\begin{equation}
\frac{\rho_{X}}{\rho_{H}}|_{\odot} =
\frac{n_{X}}{n_{H}}|_{\odot}\ M_{X},
\end{equation}
and with equation \ref{eqn:AX1}, we get
\begin{equation}
\frac{\rho_{X}}{\rho_{H}}|_{\odot} = A_{X}\ M_{X}.
\end{equation}
Equation \ref{eqn:rhox1} then becomes
\begin{equation}
\rho_{X} = Z\ \rho_{H}\ A_{X}\ M_{X}.
\end{equation}
Finally, if we don't have $\rho_{H}$, we use $\chi$ to get
\begin{equation}
\rho_{X} = Z\ \chi\ \rho\ A_{X}\ M_{X}.
\end{equation}
This gives us
\begin{equation}
n_{X} = \frac{Z \chi\ \rho\ A_{X}}{m_{H}}
\end{equation}
and
\begin{equation}
n_{X_i} = \frac{f_i Z \chi\ \rho\ A_{X}}{m_{H}}.
\end{equation}
\end{appendices}

\bibliographystyle{apj}
%\bibliography{trident.bib}

\begin{thebibliography}{}
\expandafter\ifx\csname natexlab\endcsname\relax\def\natexlab#1{#1}\fi

\bibitem[{{Allende Prieto} {et~al.}(2001){Allende Prieto}, {Lambert}, \&
  {Asplund}}]{allende01}
{Allende Prieto}, C., {Lambert}, D.~L., \& {Asplund}, M. 2001, \apjl, 556, L63

\bibitem[{{Allende Prieto} {et~al.}(2002){Allende Prieto}, {Lambert}, \&
  {Asplund}}]{allende02}
---. 2002, \apjl, 573, L137

\bibitem[{{Armstrong}(1967)}]{armstrong67}
{Armstrong}, B.~H. 1967, Journal of Quantitative Spectroscopy and Radiative
  Transfer, 7, 85

\bibitem[{{Astropy Collaboration} {et~al.}(2013){Astropy Collaboration},
  {Robitaille}, {Tollerud}, {Greenfield}, {Droettboom}, {Bray}, {Aldcroft},
  {Davis}, {Ginsburg}, {Price-Whelan}, {Kerzendorf}, {Conley}, {Crighton},
  {Barbary}, {Muna}, {Ferguson}, {Grollier}, {Parikh}, {Nair}, {Unther},
  {Deil}, {Woillez}, {Conseil}, {Kramer}, {Turner}, {Singer}, {Fox}, {Weaver},
  {Zabalza}, {Edwards}, {Azalee Bostroem}, {Burke}, {Casey}, {Crawford},
  {Dencheva}, {Ely}, {Jenness}, {Labrie}, {Lim}, {Pierfederici}, {Pontzen},
  {Ptak}, {Refsdal}, {Servillat}, \& {Streicher}}]{astropy}
{Astropy Collaboration}, {Robitaille}, T.~P., {Tollerud}, E.~J., {et~al.} 2013,
  \aap, 558, A33

\bibitem[{Behroozi {et~al.}(2012)Behroozi, Wechsler, \& Wu}]{behroozi12}
Behroozi, P.~S., Wechsler, R.~H., \& Wu, H.-Y. 2012, The Astrophysical Journal,
  762, 109

\bibitem[{{Bird} {et~al.}(2015){Bird}, {Haehnelt}, {Neeleman}, {Genel},
  {Vogelsberger}, \& {Hernquist}}]{bird15}
{Bird}, S., {Haehnelt}, M., {Neeleman}, M., {et~al.} 2015, \mnras, 447, 1834

\bibitem[{{Bordoloi} {et~al.}(2014){Bordoloi}, {Tumlinson}, {Werk},
  {Oppenheimer}, {Peeples}, {Prochaska}, {Tripp}, {Katz}, {Dav{\'e}}, {Fox},
  {Thom}, {Ford}, {Weinberg}, {Burchett}, \& {Kollmeier}}]{bordoloi14}
{Bordoloi}, R., {Tumlinson}, J., {Werk}, J.~K., {et~al.} 2014, \apj, 796, 136

\bibitem[{{Bryan} {et~al.}(2014){Bryan}, {Norman}, {O'Shea}, {Abel}, {Wise},
  {Turk}, {Reynolds}, {Collins}, {Wang}, {Skillman}, {Smith}, {Harkness},
  {Bordner}, {Kim}, {Kuhlen}, {Xu}, {Goldbaum}, {Hummels}, {Kritsuk}, {Tasker},
  {Skory}, {Simpson}, {Hahn}, {Oishi}, {So}, {Zhao}, {Cen}, {Li}, \& {Enzo
  Collaboration}}]{bryan14}
{Bryan}, G.~L., {Norman}, M.~L., {O'Shea}, B.~W., {et~al.} 2014, \apjs, 211, 19

\bibitem[{{Cen} {et~al.}(1994){Cen}, {Bahcall}, \& {Gramann}}]{cen94}
{Cen}, R., {Bahcall}, N.~A., \& {Gramann}, M. 1994, \apjl, 437, L51

\bibitem[{{Cen} \& {Fang}(2006)}]{cen06}
{Cen}, R., \& {Fang}, T. 2006, \apj, 650, 573

\bibitem[{{Chen} {et~al.}(2010){Chen}, {Helsby}, {Gauthier}, {Shectman},
  {Thompson}, \& {Tinker}}]{chen10}
{Chen}, H.-W., {Helsby}, J.~E., {Gauthier}, J.-R., {et~al.} 2010, \apj, 714,
  1521

\bibitem[{{Churchill} {et~al.}(2015){Churchill}, {Vander Vliet},
  {Trujillo-Gomez}, {Kacprzak}, \& {Klypin}}]{churchill15}
{Churchill}, C.~W., {Vander Vliet}, J.~R., {Trujillo-Gomez}, S., {Kacprzak},
  G.~G., \& {Klypin}, A. 2015, \apj, 802, 10

\bibitem[{{Conroy} {et~al.}(2009){Conroy}, {Gunn}, \& {White}}]{conroy09}
{Conroy}, C., {Gunn}, J.~E., \& {White}, M. 2009, \apj, 699, 486

\bibitem[{{Dalcin} {et~al.}(2005){Dalcin}, {Paz}, \& {Storti}}]{mpi4py}
{Dalcin}, L., {Paz}, R., \& {Storti}, M. 2005, Journal of Parallel and
  Distributed Computing, 65, 1108

\bibitem[{Danforth {et~al.}(2016)Danforth, Keeney, Tilton, Shull, Stocke,
  Stevans, Pieri, Savage, France, Syphers, Smith, Green, Froning, Penton, \&
  Osterman}]{danforth16}
Danforth, C.~W., Keeney, B.~A., Tilton, E.~M., {et~al.} 2016, {{ApJ}}, 817, 111

\bibitem[{{Dav{\'e}} {et~al.}(1997){Dav{\'e}}, {Hernquist}, {Weinberg}, \&
  {Katz}}]{dave97}
{Dav{\'e}}, R., {Hernquist}, L., {Weinberg}, D.~H., \& {Katz}, N. 1997, \apj,
  477, 21

\bibitem[{{Dav{\'e}} {et~al.}(2001){Dav{\'e}}, {Cen}, {Ostriker}, {Bryan},
  {Hernquist}, {Katz}, {Weinberg}, {Norman}, \& {O'Shea}}]{dave01}
{Dav{\'e}}, R., {Cen}, R., {Ostriker}, J.~P., {et~al.} 2001, \apj, 552, 473

\bibitem[{{Dullemond}(2012)}]{dullemond12}
{Dullemond}, C.~P. 2012, {RADMC-3D: A multi-purpose radiative transfer tool},
  Astrophysics Source Code Library, ascl:1202.015

\bibitem[{Egan {et~al.}(2014)Egan, Smith, O{\textquoteright}Shea, \&
  Shull}]{egan14}
Egan, H., Smith, B.~D., O{\textquoteright}Shea, B.~W., \& Shull, J.~M. 2014,
  The Astrophysical Journal, 791, 64

\bibitem[{{Faucher-Gigu{\`e}re} \& {Kere{\v s}}(2011)}]{CAFG11}
{Faucher-Gigu{\`e}re}, C.-A., \& {Kere{\v s}}, D. 2011, \mnras, 412, L118

\bibitem[{{Faucher-Gigu{\`e}re} {et~al.}(2009){Faucher-Gigu{\`e}re}, {Lidz},
  {Zaldarriaga}, \& {Hernquist}}]{CAFG09}
{Faucher-Gigu{\`e}re}, C.-A., {Lidz}, A., {Zaldarriaga}, M., \& {Hernquist}, L.
  2009, \apj, 703, 1416

\bibitem[{{Ferland} {et~al.}(1998){Ferland}, {Korista}, {Verner}, {Ferguson},
  {Kingdon}, \& {Verner}}]{ferland98}
{Ferland}, G.~J., {Korista}, K.~T., {Verner}, D.~A., {et~al.} 1998, \pasp, 110,
  761

\bibitem[{{Ferland} {et~al.}(2013){Ferland}, {Porter}, {van Hoof}, {Williams},
  {Abel}, {Lykins}, {Shaw}, {Henney}, \& {Stancil}}]{ferland13}
{Ferland}, G.~J., {Porter}, R.~L., {van Hoof}, P.~A.~M., {et~al.} 2013, Revista
  Mexicana de Astronomia y Astrofisica, 49, 137

\bibitem[{{Fielding} {et~al.}(2016){Fielding}, {Quataert}, {McCourt}, \&
  {Thompson}}]{fielding16}
{Fielding}, D., {Quataert}, E., {McCourt}, M., \& {Thompson}, T.~A. 2016, ArXiv
  e-prints, arXiv:1606.06734

\bibitem[{Forum(1994)}]{mpi}
Forum, M.~P. 1994, MPI: A Message-Passing Interface Standard, Tech. rep.,
  Knoxville, TN, USA

\bibitem[{{Fryxell} {et~al.}(2000){Fryxell}, {Olson}, {Ricker}, {Timmes},
  {Zingale}, {Lamb}, {MacNeice}, {Rosner}, {Truran}, \& {Tufo}}]{fryxell00}
{Fryxell}, B., {Olson}, K., {Ricker}, P., {et~al.} 2000, \apjs, 131, 273

\bibitem[{{Grevesse} \& {Sauval}(1998)}]{grevesse98}
{Grevesse}, N., \& {Sauval}, A.~J. 1998, \ssr, 85, 161

\bibitem[{{Guedes} {et~al.}(2011){Guedes}, {Callegari}, {Madau}, \&
  {Mayer}}]{guedes11}
{Guedes}, J., {Callegari}, S., {Madau}, P., \& {Mayer}, L. 2011, \apj, 742, 76

\bibitem[{{Haardt} \& {Madau}(2012)}]{haardt12}
{Haardt}, F., \& {Madau}, P. 2012, \apj, 746, 125

\bibitem[{{Hernquist} {et~al.}(1996){Hernquist}, {Katz}, {Weinberg}, \&
  {Miralda-Escud{\'e}}}]{hernquist96}
{Hernquist}, L., {Katz}, N., {Weinberg}, D.~H., \& {Miralda-Escud{\'e}}, J.
  1996, \apjl, 457, L51

\bibitem[{Hill(2016)}]{hill16}
Hill, C. 2016, Learning Scientific Programming with Python (Cambridge
  University Press)

\bibitem[{{Hogg}(1999)}]{hogg99}
{Hogg}, D.~W. 1999, ArXiv Astrophysics e-prints, astro-ph/9905116

\bibitem[{{Holweger}(2001)}]{holweger01}
{Holweger}, H. 2001, in American Institute of Physics Conference Series, Vol.
  598, Joint SOHO/ACE workshop ``Solar and Galactic Composition'', ed. R.~F.
  {Wimmer-Schweingruber}, 23--30

\bibitem[{{Hopkins}(2015)}]{hopkins15}
{Hopkins}, P.~F. 2015, \mnras, 450, 53

\bibitem[{{Hopkins} {et~al.}(2014{\natexlab{a}}){Hopkins}, {Kere{\v s}},
  {O{\~n}orbe}, {Faucher-Gigu{\`e}re}, {Quataert}, {Murray}, \&
  {Bullock}}]{hopkins14}
{Hopkins}, P.~F., {Kere{\v s}}, D., {O{\~n}orbe}, J., {et~al.}
  2014{\natexlab{a}}, \mnras, 445, 581

\bibitem[{{Hopkins} {et~al.}(2014{\natexlab{b}}){Hopkins}, {Kere{\v s}},
  {O{\~n}orbe}, {Faucher-Gigu{\`e}re}, {Quataert}, {Murray}, \&
  {Bullock}}]{hopkins14a}
---. 2014{\natexlab{b}}, \mnras, 445, 581

\bibitem[{{Hummels} \& {Bryan}(2012)}]{hummels12}
{Hummels}, C.~B., \& {Bryan}, G.~L. 2012, \apj, 749, 140

\bibitem[{{Hummels} {et~al.}(2013){Hummels}, {Bryan}, {Smith}, \&
  {Turk}}]{hummels13}
{Hummels}, C.~B., {Bryan}, G.~L., {Smith}, B.~D., \& {Turk}, M.~J. 2013,
  \mnras, 430, 1548

\bibitem[{Hunter(2007)}]{matplotlib}
Hunter, J.~D. 2007, Computing In Science \& Engineering, 9, 90

\bibitem[{{Husser} {et~al.}(2013){Husser}, {Wende-von Berg}, {Dreizler},
  {Homeier}, {Reiners}, {Barman}, \& {Hauschildt}}]{husser13}
{Husser}, T.-O., {Wende-von Berg}, S., {Dreizler}, S., {et~al.} 2013, \aap,
  553, A6

\bibitem[{{Johnson} {et~al.}(2015){Johnson}, {Chen}, \& {Mulchaey}}]{johnson15}
{Johnson}, S.~D., {Chen}, H.-W., \& {Mulchaey}, J.~S. 2015, \mnras, 452, 2553

\bibitem[{{Jonsson}(2006)}]{jonsson06}
{Jonsson}, P. 2006, \mnras, 372, 2

\bibitem[{{Kacprzak} {et~al.}(2015){Kacprzak}, {Churchill}, {Murphy}, \&
  {Cooke}}]{kacprzak15}
{Kacprzak}, G.~G., {Churchill}, C.~W., {Murphy}, M.~T., \& {Cooke}, J. 2015,
  \mnras, 446, 2861

\bibitem[{Kim {et~al.}(2014)Kim, Abel, Agertz, Bryan, Ceverino, Christensen,
  Conroy, Dekel, Gnedin, Goldbaum, Guedes, Hahn, Hobbs, Hopkins, Hummels,
  Iannuzzi, Kere{\v s}, Klypin, Kravtsov, Krumholz, Kuhlen, Leitner, Madau,
  Mayer, Moody, Nagamine, Norman, Onorbe, O'Shea, Pillepich, Primack, Quinn,
  Read, Robertson, Rocha, Rudd, Shen, Smith, Szalay, Teyssier, Thompson,
  Todoroki, Turk, Wadsley, Wise, Zolotov, \& AGORA~Collaboration29}]{kim14}
Kim, J.-h., Abel, T., Agertz, O., {et~al.} 2014, The Astrophysical Journal
  Supplement, 210, 14

\bibitem[{{Kim} {et~al.}(2016){Kim}, {Agertz}, {Teyssier}, {Butler},
  {Ceverino}, {Choi}, {Feldmann}, {Keller}, {Lupi}, {Quinn}, {Revaz},
  {Wallace}, {Gnedin}, {Leitner}, {Shen}, {Smith}, {Thompson}, {Turk}, {Abel},
  {Arraki}, {Benincasa}, {Chakrabarti}, {DeGraf}, {Dekel}, {Goldbaum},
  {Hopkins}, {Hummels}, {Klypin}, {Li}, {Madau}, {Mandelker}, {Mayer},
  {Nagamine}, {Nickerson}, {O'Shea}, {Primack}, {Roca-F{\`a}brega}, {Semenov},
  {Shimizu}, {Simpson}, {Todoroki}, {Wadsley}, {Wise}, \& {for the AGORA
  Collaboration}}]{kim16}
{Kim}, J.-h., {Agertz}, O., {Teyssier}, R., {et~al.} 2016, ArXiv e-prints,
  arXiv:1610.03066

\bibitem[{{Kravtsov}(1999)}]{kravtsov99}
{Kravtsov}, A.~V. 1999, PhD thesis, New Mexico State University

\bibitem[{{Kravtsov} {et~al.}(1997){Kravtsov}, {Klypin}, \&
  {Khokhlov}}]{kravtsov97}
{Kravtsov}, A.~V., {Klypin}, A.~A., \& {Khokhlov}, A.~M. 1997, \apjs, 111, 73

\bibitem[{{Kurucz}(1979)}]{kurucz79}
{Kurucz}, R.~L. 1979, \apjs, 40, 1

\bibitem[{{Lehner}(2016)}]{lehner16}
{Lehner}, N. 2016, ArXiv e-prints, arXiv:1612.00458

\bibitem[{{Leitherer} {et~al.}(1999){Leitherer}, {Schaerer}, {Goldader},
  {Delgado}, {Robert}, {Kune}, {de Mello}, {Devost}, \&
  {Heckman}}]{leitherer99}
{Leitherer}, C., {Schaerer}, D., {Goldader}, J.~D., {et~al.} 1999, \apjs, 123,
  3

\bibitem[{{Liang} \& {Chen}(2014)}]{liang14}
{Liang}, C.~J., \& {Chen}, H.-W. 2014, \mnras, 445, 2061

\bibitem[{{Liang} {et~al.}(2016){Liang}, {Kravtsov}, \& {Agertz}}]{liang16}
{Liang}, C.~J., {Kravtsov}, A.~V., \& {Agertz}, O. 2016, \mnras, 458, 1164

\bibitem[{{McQuinn}(2016)}]{mcquinn16}
{McQuinn}, M. 2016, \araa, 54, 313

\bibitem[{{Miralda-Escud{\'e}} {et~al.}(1996){Miralda-Escud{\'e}}, {Cen},
  {Ostriker}, \& {Rauch}}]{miralda96}
{Miralda-Escud{\'e}}, J., {Cen}, R., {Ostriker}, J.~P., \& {Rauch}, M. 1996,
  \apj, 471, 582

\bibitem[{{Nenkova} {et~al.}(2000){Nenkova}, {Ivezi{\'c}}, \&
  {Elitzur}}]{nenkova00}
{Nenkova}, M., {Ivezi{\'c}}, {\v Z}., \& {Elitzur}, M. 2000, Thermal Emission
  Spectroscopy and Analysis of Dust, Disks, and Regoliths, 196, 77

\bibitem[{{Nielsen} {et~al.}(2013){Nielsen}, {Churchill}, {Kacprzak}, \&
  {Murphy}}]{nielsen13}
{Nielsen}, N.~M., {Churchill}, C.~W., {Kacprzak}, G.~G., \& {Murphy}, M.~T.
  2013, \apj, 776, 114

\bibitem[{{Oppenheimer} \& {Dav{\'e}}(2006)}]{oppenheimer06}
{Oppenheimer}, B.~D., \& {Dav{\'e}}, R. 2006, \mnras, 373, 1265

\bibitem[{{Oppenheimer} {et~al.}(2016){Oppenheimer}, {Crain}, {Schaye},
  {Rahmati}, {Richings}, {Trayford}, {Tumlinson}, {Bower}, {Schaller}, \&
  {Theuns}}]{oppenheimer16}
{Oppenheimer}, B.~D., {Crain}, R.~A., {Schaye}, J., {et~al.} 2016, \mnras, 460,
  2157

\bibitem[{{Peebles}(1993)}]{peebles93}
{Peebles}, P.~J.~E. 1993, {Principles of Physical Cosmology}

\bibitem[{Peeples(2014)}]{peeples14}
Peeples, M. 2014, HST Proposal ID {\#}13919. Cycle 22, 13919

\bibitem[{Peeples {et~al.}(2013)Peeples, Werk, Tumlinson, Oppenheimer,
  Prochaska, Katz, \& Weinberg}]{peeples13}
Peeples, M.~S., Werk, J.~K., Tumlinson, J., {et~al.} 2013, arXiv.org, 54

\bibitem[{Poppe \& Wijers(1990)}]{poppe90}
Poppe, G. P.~M., \& Wijers, C. M.~J. 1990, ACM Trans. Math. Softw., 16, 38

\bibitem[{{Prochaska} {et~al.}(2011){Prochaska}, {Weiner}, {Chen}, {Mulchaey},
  \& {Cooksey}}]{prochaska11}
{Prochaska}, J.~X., {Weiner}, B., {Chen}, H.-W., {Mulchaey}, J., \& {Cooksey},
  K. 2011, \apj, 740, 91

\bibitem[{{Rahmati} {et~al.}(2013){Rahmati}, {Schaye}, {Pawlik}, \&
  {Raicevic}}]{rahmati13}
{Rahmati}, A., {Schaye}, J., {Pawlik}, A.~H., \& {Raicevic}, M. 2013, \mnras,
  431, 2261

\bibitem[{{Robitaille}(2011)}]{robitaille11}
{Robitaille}, T.~P. 2011, \aap, 536, A79

\bibitem[{{Rubin}(2016)}]{rubin16}
{Rubin}, K.~H.~R. 2016, ArXiv e-prints, arXiv:1612.00805

\bibitem[{{Rubin} {et~al.}(2014){Rubin}, {Prochaska}, {Koo}, {Phillips},
  {Martin}, \& {Winstrom}}]{rubin14}
{Rubin}, K.~H.~R., {Prochaska}, J.~X., {Koo}, D.~C., {et~al.} 2014, \apj, 794,
  156

\bibitem[{{Rudd} {et~al.}(2008){Rudd}, {Zentner}, \& {Kravtsov}}]{rudd08}
{Rudd}, D.~H., {Zentner}, A.~R., \& {Kravtsov}, A.~V. 2008, \apj, 672, 19

\bibitem[{{Rudie} {et~al.}(2012){Rudie}, {Steidel}, {Trainor}, {Rakic},
  {Bogosavljevi{\'c}}, {Pettini}, {Reddy}, {Shapley}, {Erb}, \&
  {Law}}]{rudie12}
{Rudie}, G.~C., {Steidel}, C.~C., {Trainor}, R.~F., {et~al.} 2012, \apj, 750,
  67

\bibitem[{{Rybicki} \& {Lightman}(1979)}]{rybicki79}
{Rybicki}, G.~B., \& {Lightman}, A.~P. 1979, {Radiative processes in
  astrophysics}

\bibitem[{{Schaye} {et~al.}(2003){Schaye}, {Aguirre}, {Kim}, {Theuns}, {Rauch},
  \& {Sargent}}]{schaye03}
{Schaye}, J., {Aguirre}, A., {Kim}, T.-S., {et~al.} 2003, \apj, 596, 768

\bibitem[{{Schaye} {et~al.}(2015){Schaye}, {Crain}, {Bower}, {Furlong},
  {Schaller}, {Theuns}, {Dalla Vecchia}, {Frenk}, {McCarthy}, {Helly},
  {Jenkins}, {Rosas-Guevara}, {White}, {Baes}, {Booth}, {Camps}, {Navarro},
  {Qu}, {Rahmati}, {Sawala}, {Thomas}, \& {Trayford}}]{schaye15}
{Schaye}, J., {Crain}, R.~A., {Bower}, R.~G., {et~al.} 2015, \mnras, 446, 521

\bibitem[{{Shen} {et~al.}(2013){Shen}, {Madau}, {Guedes}, {Mayer}, {Prochaska},
  \& {Wadsley}}]{shen13}
{Shen}, S., {Madau}, P., {Guedes}, J., {et~al.} 2013, \apj, 765, 89

\bibitem[{{Sipocz}(2016)}]{sipocz16}
{Sipocz}, B. 2016, in Python in Astronomy 2016, 34

\bibitem[{{Smith} {et~al.}(2008){Smith}, {Sigurdsson}, \& {Abel}}]{smith08}
{Smith}, B., {Sigurdsson}, S., \& {Abel}, T. 2008, \mnras, 385, 1443

\bibitem[{{Smith} {et~al.}(2011){Smith}, {Hallman}, {Shull}, \&
  {O'Shea}}]{smith11}
{Smith}, B.~D., {Hallman}, E.~J., {Shull}, J.~M., \& {O'Shea}, B.~W. 2011,
  \apj, 731, 6

\bibitem[{Smith {et~al.}(2016)Smith, Bryan, Glover, Goldbaum, Turk, Regan,
  Wise, Schive, Abel, Emerick, O{\textquoteright}Shea, Anninos, Hummels, \&
  Khochfar}]{smith16}
Smith, B.~D., Bryan, G.~L., Glover, S. C.~O., {et~al.} 2016, arXiv.org,
  arXiv:1610.09591

\bibitem[{{Springel}(2005)}]{springel05}
{Springel}, V. 2005, \mnras, 364, 1105

\bibitem[{{Springel}(2010)}]{springel10}
---. 2010, \mnras, 401, 791

\bibitem[{{Springel} {et~al.}(2001){Springel}, {Yoshida}, \&
  {White}}]{springel01}
{Springel}, V., {Yoshida}, N., \& {White}, S.~D.~M. 2001, New Astronomy, 6, 79

\bibitem[{{Steidel} {et~al.}(2010){Steidel}, {Erb}, {Shapley}, {Pettini},
  {Reddy}, {Bogosavljevi{\'c}}, {Rudie}, \& {Rakic}}]{steidel10}
{Steidel}, C.~C., {Erb}, D.~K., {Shapley}, A.~E., {et~al.} 2010, \apj, 717, 289

\bibitem[{{Stinson} {et~al.}(2006){Stinson}, {Seth}, {Katz}, {Wadsley},
  {Governato}, \& {Quinn}}]{stinson06}
{Stinson}, G., {Seth}, A., {Katz}, N., {et~al.} 2006, \mnras, 373, 1074

\bibitem[{{Stone} {et~al.}(2008){Stone}, {Gardiner}, {Teuben}, {Hawley}, \&
  {Simon}}]{stone08}
{Stone}, J.~M., {Gardiner}, T.~A., {Teuben}, P., {Hawley}, J.~F., \& {Simon},
  J.~B. 2008, \apjs, 178, 137

\bibitem[{Telfer {et~al.}(2002)Telfer, Zheng, Kriss, \& Davidsen}]{telfer02}
Telfer, R.~C., Zheng, W., Kriss, G.~A., \& Davidsen, A.~F. 2002, {ApJ}, 565,
  773

\bibitem[{{Teyssier}(2002)}]{teyssier02}
{Teyssier}, R. 2002, \aap, 385, 337

\bibitem[{{The HDF Group}(1997--)}]{hdf5}
{The HDF Group}. 1997--, {Hierarchical Data Format, version 5}, /HDF5/

\bibitem[{{Tumlinson} {et~al.}(2013){Tumlinson}, {Thom}, {Werk}, {Prochaska},
  {Tripp}, {Katz}, {Dav{\'e}}, {Oppenheimer}, {Meiring}, {Ford}, {O'Meara},
  {Peeples}, {Sembach}, \& {Weinberg}}]{tumlinson13}
{Tumlinson}, J., {Thom}, C., {Werk}, J.~K., {et~al.} 2013, \apj, 777, 59

\bibitem[{{Turk} {et~al.}(2011){Turk}, {Smith}, {Oishi}, {Skory}, {Skillman},
  {Abel}, \& {Norman}}]{turk11}
{Turk}, M.~J., {Smith}, B.~D., {Oishi}, J.~S., {et~al.} 2011, \apjs, 192, 9

\bibitem[{{Turner} {et~al.}(2015){Turner}, {Schaye}, {Steidel}, {Rudie}, \&
  {Strom}}]{turner15}
{Turner}, M.~L., {Schaye}, J., {Steidel}, C.~C., {Rudie}, G.~C., \& {Strom},
  A.~L. 2015, \mnras, 450, 2067

\bibitem[{van~der Walt {et~al.}(2011)van~der Walt, Colbert, \&
  Varoquaux}]{numpy}
van~der Walt, S., Colbert, S.~C., \& Varoquaux, G. 2011, Computing in Science
  Engineering, 13, 22

\bibitem[{{Vogelsberger} {et~al.}(2014){Vogelsberger}, {Genel}, {Springel},
  {Torrey}, {Sijacki}, {Xu}, {Snyder}, {Nelson}, \&
  {Hernquist}}]{vogelsberger14}
{Vogelsberger}, M., {Genel}, S., {Springel}, V., {et~al.} 2014, \mnras, 444,
  1518

\bibitem[{{Wadsley} {et~al.}(2004){Wadsley}, {Stadel}, \& {Quinn}}]{wadsley04}
{Wadsley}, J.~W., {Stadel}, J., \& {Quinn}, T. 2004, New Astronomy, 9, 137

\bibitem[{{Zhang} {et~al.}(1995){Zhang}, {Anninos}, \& {Norman}}]{zhang95}
{Zhang}, Y., {Anninos}, P., \& {Norman}, M.~L. 1995, \apjl, 453, L57

\end{thebibliography}

\end{document}